\documentclass{lmcs}

\pdfoutput=1


\usepackage{hyperref}
\usepackage{upgreek}
\usepackage{physics}
\usepackage{stmaryrd}
\usepackage{mathtools}
\usepackage{wasysym}
\usepackage{cleveref}

\newcommand{\emptyword}{\upvarepsilon}

\newcommand{\sub}{\theta}

\newcommand{\lang}{\mathcal{L}}

\newcommand{\natnums}{\mathbb{N}}

\newcommand{\fcdatalogprogram}{P}

\newcommand{\anotherfcdatalogprogram}{Q}

\newcommand{\fcdatalogrule}{\rho}

\newcommand{\anotherfcdatalogrule}{\chi}

\newcommand{\fcdatalogruleset}{\Phi}

\newcommand{\atom}{\varphi}

\newcommand{\inputsymbol}{\mathfrak{u}}

\newcommand{\projection}{\pi}

\newcommand{\join}{\Bowtie}

\newcommand{\stringeq}{\zeta^{=}}

\newcommand{\terminals}{\Upsigma}

\newcommand{\vars}{\Xi}

\newcommand{\relationsymbols}{\mathcal{R}}

\newcommand{\relationsymbol}{R}

\newcommand{\pattern}{\alpha}

\newcommand{\bigo}{\mathcal{O}}

\newcommand{\cq}{\mathsf{CQ}}

\newcommand{\theoryofconcatenation}{\mathsf{C}}

\newcommand{\fo}{\mathsf{FO}}

\newcommand{\mso}{\mathsf{MSO}}

\newcommand{\fc}{\mathsf{FC}}

\newcommand{\fcreg}{\mathsf{FC[REG]}}

\newcommand{\epfc}{\mathsf{EP}$-$\mathsf{FC}}

\newcommand{\epfcreg}{\mathsf{EP}$-$\mathsf{FC[REG]}}

\newcommand{\olla}{\mathsf{OLLA}}

\newcommand{\dolla}{\mathsf{DOLLA}}

\newcommand{\sd}{\mathsf{SD}}

\newcommand{\sddolla}{\mathsf{SD}$-${\mathsf{DOLLA}}}

\newcommand{\sddollaplus}{\mathsf{SD}$-${\mathsf{DOLLA+}}}

\newcommand{\dollaplus}{\mathsf{DOLLA+}}

\newcommand{\fcdatalog}{\mathsf{FC}$-$\mathsf{Datalog}}

\newcommand{\fccq}{\mathsf{FC}$-$\mathsf{CQ}}

\newcommand{\hefs}{\mathsf{HEFS}}

\newcommand{\datalog}{\mathsf{Datalog}}

\newcommand{\rgxlog}{\mathsf{RGXlog}}

\newcommand{\spannerlogrgx}{\mathsf{Spannerlog}\langle\mathsf{RGX}\rangle}

\newcommand{\p}{\mathsf{P}}

\newcommand{\np}{\mathsf{NP}}

\newcommand{\pspace}{\mathsf{PSPACE}}

\newcommand{\npspace}{\mathsf{NPSPACE}}

\newcommand{\logspace}{\mathsf{LOGSPACE}}

\newcommand{\nlogspace}{\mathsf{NLOGSPACE}}

\newcommand{\exptime}{\mathsf{EXP}}

\newcommand{\dtmfa}{\mathsf{DTMFA}}

\newcommand{\rcg}{\mathsf{RCG}}

\newcommand{\prcg}{\mathsf{PRCG}}

\newcommand{\drx}{\mathsf{DRX}}

\newcommand{\inputdatabase}{D}

\newcommand{\deterministicregex}{\gamma}

\newcommand{\ans}{\text{Ans}}

\newcommand{\ar}{\text{ar}}

\newcommand{\Var}{\text{Var}}

\newcommand{\signature}{\sigma}

\newcommand{\structure}{\mathfrak{A}}

\newcommand{\we}{\mathsf{pe}}

\newcommand{\topvar}{\mathsf{top}}

\newcommand{\bottomvar}{\mathsf{bottom}}

\newcommand{\factor}{\sqsubseteq}

\newcommand{\free}{\mathsf{free}}
\newcommand{\accept}{\mathsf{Accept}}
\newcommand{\reject}{\mathsf{Reject}}
\newcommand{\probl}{\mathsf{Prob}}

\DeclarePairedDelimiter{\ceil}{\lceil}{\rceil}

\newcommand{\rc}{,\hspace{0.5em}}

\newcommand{\smallrc}{,\hspace{0.3em}}

\newcommand{\textrc}{,\hspace{0.2em}}

\newcommand{\inlineextraspace}{\hspace{0.3em}}

\newcommand*{\changetwo}{\textcolor{black}}
\newcommand*{\changethree}{\textcolor{black}}
\newcommand*{\changefour}{\textcolor{black}}

\newcommand*{\changetwofour}{\textcolor{black}}

\newcommand*{\changeall}{\textcolor{black}}

\newcommand*{\changeother}{\textcolor{black}}
	
\begin{document}
		
	\title[FC-Datalog as a Framework for Efficient String Querying]{FC-Datalog as a Framework for Efficient String Querying}
	\titlecomment{This is an extended version of an article published at ICDT 2025 \cite{ICDT25}}
	\thanks{This work was supported by EPSRC grant EP/T033762/1.}
				
	\author[O.~M.~Bell]{Owen M. Bell\lmcsorcid{0009-0007-8262-2813}}
	\author[J.~D.~Day]{Joel D. Day\lmcsorcid{0000-0002-3660-7766}}
	\author[D.~D.~Freydenberger]{Dominik~D.~Freydenberger\lmcsorcid{0000-0001-5088-0067}}
		
	\address{Loughborough University, UK}
	\email{o.m.bell.tcs@gmail.com, j.day@lboro.ac.uk, d.d.freydenberger@lboro.ac.uk}  
				
	\begin{abstract}
		\noindent 
		Core spanners are a class of document spanners that capture the core functionality of IBM's AQL. $\fc$~is a logic on strings built around word equations that when extended with constraints for regular languages can be seen as a logic for core spanners. 
		\changetwo{The recently introduced} $\fcdatalog$ extends $\fc$ with recursion, which allows us to define recursive relations for core spanners. Additionally, as $\fcdatalog$ captures~$\p$, it is also a tractable version of $\datalog$ on strings. This presents an opportunity for optimization.
		
		We \changetwo{propose} a series of $\fcdatalog$ fragments with desirable properties in terms of complexity of model checking, expressive power, and efficiency of checking membership in the fragment. This leads to a \changetwo{range} of fragments that all capture $\logspace$, which we further restrict to obtain linear combined complexity. 
		This gives us a framework to tailor fragments for particular applications. To showcase this, we simulate deterministic regex in a tailored fragment of $\fcdatalog$.
	\end{abstract}
		
	\maketitle
				
	\section{Introduction}
		\label{sec:introduction}
		As a vast amount of valuable information is stored in unstructured textual data, the operation of extracting structured information from such data is crucial. This operation is the classical task known as \emph{Information Extraction} (IE), and has applications from healthcare~(see~e.g.~\cite{IEHealthcare}) to social media analytics (see e.g. \cite{IESocialMedia}). The rule-based approach to IE can be understood as querying a text in the same way as one queries a relational database. Here we consider a recursive model for rule-based IE called $\fcdatalog$, which is based on the logic on strings $\fc$ and the query language $\datalog$. 

		In $\fc$ and $\fcdatalog$, we reason over factors (contiguous or non-breaking subwords) of a string. $\fc$ has tight connections to the well-studied rule-based IE framework of document spanners, which reasons over intervals of positions in a string. By reasoning directly over factors, we can simplify the way we express queries, whilst maintaining the ability to express precise intervals if this is required. We now give an introduction to the two components of our model, $\fc$ and $\datalog$, as well as a detailed explanation of the connection to document spanners, before giving an introduction to $\fcdatalog$.

		\subsection*{FC}
			The \emph{theory of concatenation} (short: $\theoryofconcatenation$) is a logic on strings which has the infinite universe~$\terminals^\ast$. Introduced in~\cite{FC}, $\fc$ is a finite model version of $\theoryofconcatenation$ which has a finite universe of a single word and all of its factors (contiguous or non-breaking subwords). As a result of this restriction, $\fc$ has decidable \emph{model checking} and \emph{evaluation} (see Section 4 of \cite{FC}). $\fc$ is built on \emph{word equations}, that is equations of the form $xx \doteq yyy$, where variables $x$ and $y$ represent words over a finite alphabet $\terminals$. As a result, in $\fc$ we can reason directly over factors rather than intervals of positions as is the case for other logics on strings such as \emph{monadic second order logic} ($\mso$) over a linear order. Furthermore, in $\fc$ we can compare factors of unbounded length, which is not possible in $\mso$ (see \cite{FC} for details). 
			
			Word equations themselves are widely studied and are a natural way of expressing many typical string properties such as if the string is imprimitive or if it contains a square (see e.g.~\cite{wordEquationExpPower}), and have previously been used as a model for data management in other areas such as graph databases (see \cite{Barcelo}). Whilst here we are concerned with information extraction, $\fc$ also has applications outside this domain; in particular, $\fc$ and its various extensions, such as those defined later in the this article, represent a unifying framework for combining parsers in the context of Language-Theoretic Security (LangSec) (see \cite{LangSec} for details).

		\subsection*{Document Spanners and FC}
			Introduced in~\cite{Spanners}, \emph{document spanners} (or \emph{spanners}) are a rule-based framework for Information Extraction. Spanners were introduced to capture the core functionality of AQL, a query language used for IBM's SystemT. Informally, spanners are functions that take a text document as input and output a relation over intervals (called \emph{spans}) from the document. \changetwo{Intuitively, primitive extractors (which are commonly regular expressions with capture variables called \emph{regex formulas}) extract relations of spans, and these are then combined using relational algebra.}
			
			Many works on spanners, particularly in the area of enumeration (see e.g. \cite{constantDelay2, constantDelay, constantDelay3}), have been concerned with the subclass of \emph{regular spanners}, which are regex formulas extended with projection ($\projection$), union ($\cup$) and natural join ($\join$). However, \cite{Spanners} showed that this subclass cannot express more than the recognizable relations. The full class of spanners introduced in \cite{Spanners} are called the \emph{core spanners} as these achieve the original motivation of capturing the core functionality of AQL. Core spanners extend the regular spanners with string equality (denoted $\stringeq$), an operation necessary to perform fundamental tasks such as reading multiple occurrences of a string. This added expressivity comes at the cost of reduced efficiency (see e.g. \cite{splog, DominikMario}). Further extending the core spanners with set difference gives the class of \emph{generalized core spanners}.
			
			The logic $\fc$ has a tight connection to core spanners. Although ``pure'' $\fc$ cannot define all the regular languages (see \cite{Thompson25}), we can define $\fcreg$, the extension of $\fc$ with constraints that decide membership of regular languages. $\fcreg$ captures the expressive power of generalized core spanners, and its existential-positive fragment $\epfcreg$ captures the expressive power of core spanners. Furthermore, there are polynomial time conversions between $\fcreg$ and generalized core spanners, and $\epfcreg$ and core~spanners (see Section 5.2 of \cite{FC}). Whilst spanners reason over intervals of positions, $\fc$ reasons directly over factors. When dealing with factors, unlike with intervals of positions, the default is not to distinguish between duplicates. However, simulating different intervals containing the same factor is easily done: we can simply store in addition to the factor, the prefix preceding it. On the other hand, eliminating duplicates when working with intervals, such as in spanners, is not so easily achieved. Due to the connection between the two models, we can use $\fc$ to gain insights into spanners; previous examples of work on $\fc$ that has produced results for spanners are \cite{SplittingAtoms} for tractability \changetwo{and} \cite{EFGames} for inexpressibility. 

		\subsection*{Datalog.}
			A query language for relational databases, $\datalog$ was introduced to perform operations that were not possible in earlier database languages such as graph transitive closure (see e.g. \cite{Alice}). A $\datalog$ program has a database of prepopulated \emph{extensional} relations and a set of recursive rules that define new \emph{intensional} relations. \changethree{Semi-positive $\datalog$, which allows negation for atoms with extensional relation symbols, captures $\p$ on ordered structures (see e.g. \cite{Libkin})}. \changeall{Linear $\datalog$ permits at most one atom with an intensional relation symbol in the body of every rule, and semi-positive linear $\fcdatalog$ captures $\nlogspace$ on ordered structures (see e.g. \cite{Gottlob2003}).} A more general definition of linear $\datalog$ permits, in the body of every rule, at most one atom with an intensional relation symbol mutually recursive with the head relation symbol (see e.g \cite{Alice}). We can evaluate body atoms that have other intensional relation symbols as subroutines, and so this extended definition does not affect the complexity. Linear $\datalog$ has been investigated in the context of capturing linear recursion in SQL~(see \cite{SQL}).
			
			The combined complexity of model checking $\datalog$ is $\exptime$-\changetwo{complete}, even if the input database is empty, the universe is made up of only two elements, and the program has only a single rule (see e.g. \cite{DatalogComplexity, Gottlob2003}). For linear $\datalog$ it is $\pspace$-complete, again even with an empty input database, a two-element universe, and a single rule (see \cite{Gottlob2003}). 
			
			An earlier approach to adapting $\datalog$ for querying strings is Sequence $\datalog$ (see~\cite{Bonner}), but this has an undecidable model checking problem. Furthermore, in the spanner setting \cite{RecursiveSpanners} introduced $\rgxlog$, $\datalog$ over regex formulas. $\rgxlog$ was motivated by the SystemT developers' interest in recursion (for example, to implement context free grammars for natural language processing), and captures the complexity class $\p$. \changetwo{As introduced in \cite{Spannerlog}, $\spannerlogrgx$ generalizes the spanner and relational model and has recently been implemented in \cite{SpannerLib}. $\spannerlogrgx$ with stratified negation and restricted to string extensional relations also captures $\p$ (see Section 6 of \cite{RecursiveSpanners})}.  

		\subsection*{FC-Datalog}
			Together with $\fc$, \cite{FC} also introduced the $\fc$-analog of $\datalog$ called $\fcdatalog$, which extends \changetwo{existential-positive $\fc$ ($\epfc$) with recursion analogously to how $\datalog$ extends existential-positive $\fo$. It is worth pointing out that $\epfc$ is able to express the inequality of two strings (see e.g. Example 5.3 in \cite{splog})}. As in $\fc$, we have the finite universe of a single word and all of its factors. $\fcdatalog$ has word equations in place of the extensional relations of classical $\datalog$. That is, $\fcdatalog$ atoms are word equations or relations. In $\fcdatalog$ we adopt the \emph{fixed~point} $\datalog$ semantics.
			
			\begin{exa}
				\label{ex:fc-datalog-program-intro}
				The rules of an $\fcdatalog$ program $\fcdatalogprogram$:
				\begin{align*}
					\ans() &\shortleftarrow \inputsymbol \doteq yz\rc E(y,z);\\
					E(x, y) &\shortleftarrow x \doteq \emptyword \rc y \doteq \emptyword;\\
					E(x, y) &\shortleftarrow x \doteq \mathsf{a}u \rc y \doteq \mathsf{b}v \rc E(u,v).
				\end{align*}
				The program $\fcdatalogprogram$ defines the language $\lang(\fcdatalogprogram) \coloneqq \{ \mathsf{a}^{n}\mathsf{b}^{n} \mid n \in \natnums \}$. Informally, $\inputsymbol$ represents our input word, $\ans$ is our output relation, and a word equation holds if both sides are mapped to the same string. See Definition \ref{def:fcdatalog} for the full semantics.
			\end{exa}
			
			By Theorem 4.11 of \cite{FC}, $\fcdatalog$ captures the complexity class $\p$. When considering efficiency, we are primarily interested in model checking, which relates practically to deciding if a tuple is in a relation. \changeall{We can see $\fcdatalog$ as a language for expressing relations that can be used in spanner selections}. Because spanners reason over intervals of positions, expressing relations can become cumbersome as a relation holds such intervals, including all those that represent the same factor. In contrast, we can neatly express relations in $\fcdatalog$ as we reason directly over factors and so a relation holds the factors themselves. 
			
			\begin{exa}
				In $\fcdatalog$, we would express a relation that contains all factors that are squares with $R(x) \shortleftarrow x \doteq yy$. In core spanners, we would express a relation that contains the positions of all factors that are squares with $\pi_{x}(\zeta^{=}_{y_1, y_2} (\terminals^{\ast}x\{y_1\{\terminals^{\ast}\}y_2\{\terminals^{\ast}\}\}\terminals^{\ast}))$. See~\cite{Spanners} for the full definition of core spanners.
			\end{exa}
			
			Parallel to this, where model checking Sequence $\datalog$ is undecidable, the same problem for $\fcdatalog$ is \changetwo{in $\p$}. Hence we can also see $\fcdatalog$ as a tractable recursive query language for strings, independent of the connection to spanners. We aim to identify techniques that make recursion less expensive. We thus look to define restrictions that lead to more efficient fragments of $\fcdatalog$ which also retain other desirable properties.
			
			We focus on the model checking problem primarily through two different lenses: \emph{data complexity} and \emph{combined complexity}. In many cases for query languages it is reasonable to use data complexity as the queries are often significantly smaller than the data. 
			In text-based settings on the other hand, features  such as regular expressions can make queries large. Consequently, combined complexity remains an important consideration.
			
		\subsection*{Deterministic Regex.}
			As the regular languages are often not enough to express what is required in practice, almost all modern programming languages (such as e.g. PERL, Python, and Java) do not implement only classical \emph{regular expressions}, as introduced in~\cite{Kleene}, but \emph{regex}, regular expressions extended with \emph{back{-}references}. These are operators that match a repetition of a previously matched string, and whilst they do increase expressivity, they also lead to intractability \changetwo{of membership} (see \cite{regex}). Regex were combined with the notion of determinism in \cite{deterministicRegex} to define $\drx$, the class of \emph{deterministic regex}, a tractable class of regex with more expressive power than deterministic regular expressions. 
			
		\subsection*{Other Related Models.}
			As well as $\fcdatalog$, there also exist other related models that capture $\p$. These include positive Range~Concatenation Grammars~($\prcg$)~(see e.g.~\cite{rcgChapter}), and Hereditary~Elementary~Formal~Systems~($\hefs$)~(see e.g. \cite{introHEFS}), which do not have finite model semantics. The key difference of these models to $\fcdatalog$ is in $\fcdatalog$'s use of word equations, \changetwo{which are not present} in other formalisms and are crucial for our restrictions that lead to efficient fragments. 
			
		\subsection*{Contributions of this Paper.}
			The only previously known complexity result for $\fcdatalog$ is that it captures $\p$, and so we first show combined complexity of $\fcdatalog$ is $\exptime$-complete. We then perform an evaluation of different restrictions on $\fcdatalog$ to identify fragments that: have more efficient complexity of model checking, for both data and combined complexity; do not overly sacrifice expressive power; and have efficient checking of membership in the fragment. As the \changethree{semi-positive} linear fragment of classical $\datalog$ captures $\nlogspace$ (on ordered structures), our first restriction is to adapt linearity to $\fcdatalog$. We show that \emph{linear} $\fcdatalog$ also captures $\nlogspace$ and has $\pspace$-complete combined complexity. Our second restriction is to eliminate nondeterminism. Here, we define \emph{deterministic linear} $\fcdatalog$ which captures $\logspace$. However, checking whether a linear program is deterministic is as hard as satisfiability for word equations, a problem which is known to be $\np$-hard~(see \cite{AngluinPatterns, Koscielski}) and in nondeterministic linear space~(see \cite{Jez}).
			
			Therefore, we employ another restriction that we call \emph{one~letter~lookahead} ($\olla$) on the permitted word equations. Deterministic $\olla$ ($\dolla$) $\fcdatalog$ captures $\logspace$ and checking if an $\olla$ program is deterministic can be done in polynomial time. On the other hand, its combined complexity is still $\pspace$-complete. We thus make a final restriction that we call \emph{strictly decreasing}~($\sd$) and define $\sddolla$ $\fcdatalog$, which has linear combined complexity.
			
			We hence establish \changetwo{the endpoints of a range of fragments that all capture $\logspace$; at one end is deterministic linear $\fcdatalog$ at the other is $\dolla$ $\fcdatalog$}. We establish a trade-off in this \changetwo{range} between how rich the fragment's syntax is and how easy it is to check membership in the fragment\changetwo{, although fully mapping this range is left for future work. Furthermore, we show that we can obtain an $\fcdatalog$ fragment with linear combined complexity, namely $\sddolla$ $\fcdatalog$.} Consequently, we have \changetwo{paved the way to} design tailored fragments for particular applications.
			
			\changeall{We then explain how $\fcdatalog$ programs from our range can be viewed as generalized multi-headed two-way finite automata, which are equivalent to $\logspace$ Turing machines (see~e.g~\cite{Hartmanis72, Kozen06}). $\fcdatalog$ fragments from our range allow for more flexibility than these automata models, for example $\dollaplus$ $\fcdatalog$, tailored to simulate $\drx$, can be viewed as a generalization that permits performing nonregular string computations in the transitions}.
			
			In \cite{deterministicRegex} to check if a regex matches a word one has to construct a technically involved automata model. In contrast, we show how this task can be modelled simply in $\sddollaplus$ $\fcdatalog$. We also show how this tailored fragment allows us to conveniently and naturally write programs that are more concise. $\dollaplus$ $\fcdatalog$ permits additional deterministic components, but despite this maintains all of the desirable properties of $\dolla$ $\fcdatalog$: it captures $\logspace$, determinism can be checked in polynomial time, and its strictly decreasing variant $\sddollaplus$ $\fcdatalog$ has linear combined complexity. Finally, we show that as we can simulate $\drx$, another example of deterministic components that can be added to these tailored fragments are constraints that match $\drx$.
			
		\subsection*{Structure of this Paper.}
			Section \ref{sec:preliminaries} contains some preliminary definitions. We begin Section \ref{sec:efficient} by showing the combined complexity for unrestricted $\fcdatalog$, before moving to define our restrictions. The first, linearity, is given in Section \ref{subsec:linearity} and the second, determinism, in given in Section \ref{subsec:determinism}. This results in a fragment with efficient data complexity, but checking membership of this fragment is expensive. In Section \ref{subsec:olla} we define a further restriction, one letter lookahead, which results in a fragment that has efficiency for both data complexity and checking of membership, without sacrificing expressive power. We thus have a range of efficient fragments with this fragment as one endpoint, and the fragment defined in the previous section as the other. The remainder of Section~\ref{sec:framework} gives a final restriction, strictly decreasing, which results in efficient combined complexity, and explains how our fragments are generalizations of automata. Section \ref{sec:application} addresses the range of efficient fragments and demonstrates a tailored fragment from this range that retains all of the desirable properties we have studied, whilst permitting more convenient and natural writing of programs. We then close the paper in Section~\ref{sec:conclusions}.
			
		\subsection*{Related Version.}
			The present paper is an extended version of \cite{ICDT25}, which appeared at ICDT 2025. This version includes full proofs and the new Proposition 3.6. 
		
	\section{Preliminaries}
		\label{sec:preliminaries}
		Let $\natnums \coloneqq \{0,1,2,\ldots\}$. We denote the \emph{empty set} by $\emptyset$ and the \emph{cardinality} of a set $S$ by $\abs{S}$. Let $S \setminus S'$ be the \emph{set difference} of $S$ and $S'$, and let $\mathcal{P}(S)$ be the \emph{power set} of $S$. Let $A$ be some alphabet. For any two $u, v \in A^{\ast}$, we use $u \cdot v$, or simply $uv$, for the \emph{concatenation} of $u$ and $v$. For $u \in A^{\ast}$, let $\abs{u}$ denote its length. Let $\terminals$ be a finite alphabet that we call \emph{terminal symbols} (or just \emph{terminals}). We call any $w \in \terminals^{\ast}$ a \emph{word} (or \emph{string}) and use $\emptyword$ to denote the \emph{empty word}. A \emph{factor} of a word $w$ is a word $t$ such that there exist $u,v \in \terminals^{\ast}$ with $w = u \cdot t \cdot v$, and is denoted $t \sqsubseteq w$. In literature, a factor is sometimes called a contiguous or non-breaking subword. 

		For a \emph{tuple} $\vec{t}$, let \changetwo{its} \emph{size} $\abs{\vec{t}}$ be the number of components in $\Vec{t}$ and let $x \in \vec{t}$ denote that $x$ is a component of $\vec{t}$. We refer to tuples where $\abs{\vec{t}} = 1$ as \emph{singletons} and omit the brackets in this case. A \emph{relation} is a set of tuples of the same size, and is represented by a \emph{relation symbol}. 
		
		Let $\vars$ be a countably infinite set of \emph{variables} where $\terminals \cap \vars = \emptyset$. A \emph{pattern} is a word $\pattern \in (\terminals \cup \vars)^{\ast}$. Let $\Var(\pattern$) be the set of variables in $\pattern$. For two alphabets $A$ and $B$, a~\emph{homomorphism} is a function $g : A^{\ast} \shortrightarrow B^{\ast}$ where $g(u)$ $\cdot$ $g(v)$ = $g(u \cdot v)$ holds for all $u, v \in A^{\ast}$. A \emph{pattern~substitution} $\sub$ is a homomorphism $\sub: (\terminals \cup \vars)^{\ast} \shortrightarrow (\terminals \cup \vars)^{\ast}$ where $\sub(\mathsf{a}) = \mathsf{a}$ for all $\mathsf{a} \in \terminals$. We usually refer to a pattern substitution as simply a \emph{substitution}. We denote the image of a pattern $\pattern$ under a substitution $\sub$ by $\sub(\pattern)$. If $\sub(x) = \emptyword$ for some variable $x$, we say $\sub$ is \emph{erasing}. For a tuple $\Vec{t} = (\alpha_1, \ldots, \alpha_n)$ and a substitution $\sub$, let $\sub(\vec{t}) = (\sub(\alpha_1), \ldots, \sub(\alpha_n)) $.
		
		For two patterns $\alpha, \beta \in (\terminals \cup \vars)^{\ast}$, an equation of the form $\alpha \doteq \beta$ is called a \emph{word equation}. Under a substitution $\sub$, the word equation $\atom \coloneqq \alpha \doteq \beta$ holds if $\sub(\alpha)=\sub(\beta)$, and we then say $\sub$ is a \emph{solution} of $\atom$. If there exists a solution for $\atom$, then we say $\atom$ is \emph{satisfiable}. We say a conjunction of two word equations $\atom_1 \wedge \atom_2$ is satisfiable if there exists a substitution $\sub$ that is a solution to both $\atom_1$ and $\atom_2$. We say two word equations $\atom_1$ and $\atom_2$ \emph{contradict} each other if their conjunction is not satisfiable.

		\subsection*{FC--Datalog.}
			An $\fcdatalog$ \emph{atom} is either a pattern equation, or a so-called \emph{relation atom} of the form $R(y_{1}, \ldots, y_{\ar(R)})$ where $R$ is a relation symbol that has an arity $\ar(R) \geq 0$ and $y_i \in \vars$ for $1 \leq i \leq \ar(R)$. W.l.o.g. we can assume that for every pattern equation atom $x \doteq \pattern$, the variable $x$ does not occur in $\pattern$ because such equations reduce to trivial scenarios, as in Lemma 3.1 of \cite{Thompson24}. For a relation atom $\atom = R(y_{1}, \ldots, y_{\ar(R)})$, we define $\Var(\atom$)~=~\{$y_{1}, \ldots, y_{\ar(R)}$\}. For now, we assume each relation symbol $R$ has a corresponding relation $R$. How relations are updated is specified in the following semantics. 
			
			Let $w \in \terminals^{\ast}$ be the \emph{input word}. Then our universe is defined as $\{t \mid t \factor w\}$. That is $w$ is the word that defines the universe. In an $\fcdatalog$ program, we represent $w$ with the distinguished \emph{universe variable} $\inputsymbol \in \vars$. As our universe is dependent on $w$, whether an atom holds under a substitution $\sub$ is dependent on $w$ as well as $\sub$. We say a \emph{$w$-substitution} is a substitution $\sub$ where $\sub(\inputsymbol)=w$ and $\sub(x) \factor w$, for all $x \in \vars$. Using only $w$-substitutions ensures that the universe is restricted to $w$ and its factors. For a pattern equation atom $\atom = x \doteq \pattern$, we say $(w,\sub) \models \atom$ if $\sub(x)=\sub(\pattern)$ and we have both $\sub(x) \factor w$ and $\sub(\pattern) \factor w$. For a relation atom $\atom = R(y_{1},\ldots, y_{\ar(R)})$, we say $(w,\sub) \models \atom$ if $(\sub(y_{1}),\ldots, \sub(y_{\ar(R)})) \in R$ and $\sub(y_i) \factor w$, for $1 \leq i \leq \ar(R)$. As we only consider the finite universe setting, we need only consider $w$-substitutions, and so refer to these as simply substitutions for brevity. 
			
			\changetwo{Let $\signature$ be a relational vocabulary (see e.g. \cite{EbbinghausFlum, Libkin} for a definition of $\fo$ and vocabularies)}. A \emph{conjunctive query} is an $\fo[\signature]$ formula of the form $\fcdatalogrule(\vec{x}) \coloneqq \exists \vec{y} \colon \bigwedge_{i=1}^{m} \eta_i$, where $\eta_i \coloneqq R_{i} (\vec{t}_{i})$, each $R_{i}$ is a relation symbol in $\sigma$, each $\vec{t}_{i}$ is a $\sigma$-term, and $n \geq 1$. We usually write this as $\fcdatalogrule \coloneqq \ans(\vec{x}) \shortleftarrow \eta_1, \ldots, \eta_m$ where $\ans$ is a distinguished output relation symbol and $\vec{x}$ is the tuple of free variables. We call $\ans(\vec{x})$ the \emph{head} of $\fcdatalogrule$ and call $\eta_1, \ldots, \eta_m$ the \emph{body} of $\fcdatalogrule$. If $\ar(\ans) = 0$, then $\fcdatalogrule$ is \emph{Boolean}. We call the set of conjunctive queries $\cq$. By convention, we adopt the use of $\ans$ as the distinguished output relation symbol in $\fcdatalog$.
			
			\begin{defi}
			\label{def:fcdatalog}
			An $\fcdatalog$ \emph{program} is a 3-tuple $P \coloneqq (\inputsymbol$, $\relationsymbols$, $\fcdatalogruleset)$, where:
			\begin{itemize}
			    \item $\inputsymbol \in \vars$ is the distinguished universe variable.
			    \item $\relationsymbols$ is a finite set of relation symbols, where each $\relationsymbol \in \relationsymbols$ has an arity $\ar(R) \geq 0$, and the distinguished output relation symbol $\ans \in \relationsymbols$.
			    \item $\fcdatalogruleset$ is a finite set of \emph{rules} of the form $R(\vec{x}) \shortleftarrow \atom_{1}, \ldots, \atom_{m}$ for some $m \geq 1$, some $\relationsymbol \in \relationsymbols$, where for $1 \leq i \leq m$ every $\atom_{i}$ is an $\fcdatalog$ atom, and every $x \in \Vec{x}$ occurs in some $\text{\Var}(\atom_{i})$.
			\end{itemize}
			\end{defi}
			Each element of $\fcdatalogruleset$ can be intuitively seen as a conjunctive query over $\fcdatalog$ atoms and, as for $\cq$, when $\ar(\ans) = 0$, the program is Boolean. Also as for $\cq$, for an $\fcdatalog$ rule $R(\vec{x}) \shortleftarrow \atom_{1}, \ldots, \atom_{m}$, we say the \emph{head} of $\fcdatalogrule$ is $R(\vec{x})$ and the \emph{body} of $\fcdatalogrule$ is $\atom_{1}, \ldots, \atom_{m}$. We then call $R$ the \emph{head relation symbol} of $\fcdatalogrule$ and let $\free(\fcdatalogrule) = \{x \mid x \in \vec{x}\}$, the set of free variables in $\fcdatalogrule$. For brevity, just $\fcdatalogruleset$ is used to represent the whole tuple if $\inputsymbol$ and $\relationsymbols$ are clear from context. \changefour{For a relation symbol $R \in \relationsymbols$, let $\fcdatalogruleset_{R} \subseteq \fcdatalogruleset$ be the subset of rules with head relation symbol $R$}. For an $\fcdatalog$ rule $\fcdatalogrule$ we define $\we(\fcdatalogrule)$ as the conjunction of all the pattern equations in the body of $\fcdatalogrule$.
			
			\begin{exa}
				For the $\fcdatalog$ program $\fcdatalogprogram = (\inputsymbol, \{\ans, E\}, \fcdatalogruleset)$ where $\fcdatalogruleset$ are the rules defined in Example \ref{ex:fc-datalog-program-intro}, we have \changefour{the two subsets: $\fcdatalogruleset_{\ans} \coloneqq \{\ans() \shortleftarrow \inputsymbol \doteq yz \rc E(y,z)\}$ and $\fcdatalogruleset_{E} \coloneqq \fcdatalogruleset \setminus \fcdatalogruleset_{\ans}$}. Let $\fcdatalogrule = E(x, y) \shortleftarrow x \doteq \mathsf{a}u \smallrc y \doteq \mathsf{b}v \smallrc E(u,v)$. Then $\we(\fcdatalogrule) = x\doteq\mathsf{a}u \wedge y\doteq\mathsf{b}v$.
			\end{exa}
			
			For the $\fcdatalog$ rule $\fcdatalogrule = R(\vec{x}) \shortleftarrow \atom_{1}, \ldots, \atom_{m}$ and a $w$-substitution~$\sub$, we say $(w, \sub) \models \fcdatalogrule$ if for all $1 \leq i \leq m$ we have $(w, \sub) \models \atom_{i}$. We treat an $\fcdatalog$ program~$\fcdatalogprogram=(\inputsymbol, \relationsymbols, \fcdatalogruleset)$ as implicitly defining a vocabulary and define corresponding structures. An $\fcdatalog$ structure $\structure_{\fcdatalogprogram}$ consists of the fixed universe $\{t \mid t \factor w\}$, and an interpretation function $f^{\structure_{\fcdatalogprogram}}$ that maps every relation symbol $\relationsymbol \in \relationsymbols$ to an interpretation~$R^{\structure_{\fcdatalogprogram}}$. For convenience, we also refer to $R^{\structure_{\fcdatalogprogram}}$ as $R$.
			
			A program $\fcdatalogprogram$ and a word $w$ define a structure $\llbracket P \rrbracket (w)$ incrementally. First, all $R \in \relationsymbols$ are initialized to $\emptyset$. Then, for each rule $R(\vec{x}) \shortleftarrow \atom_{1},\ldots, \atom_{m}$, for every $w$-substitution~$\sub$ where $(w,\sub) \models \atom_{i}$ for all $1 \leq i \leq m$, add $(\sub(\vec{x}))$ to $R$. This is then repeated this until all $\relationsymbol \in \relationsymbols$ have stabilized. Then $\llbracket P \rrbracket (w)$ is the content of the $\ans$ relation. This `filling up' of relations mirrors the fixed point semantics of classical $\datalog$. This function that maps $w$ to $\ar(\ans)$-ary tuples over factors of $w$ is well-defined (thus the fixed point is unique) and the `filling up' process terminates for every given~$w$. If $\fcdatalogprogram$ is Boolean, then $\llbracket P \rrbracket (w)$ is either~$\emptyset$ or $\{()\}$. We interpret these as $\reject$ and $\accept$ respectively and use this to define the \emph{language} $\lang(\fcdatalogprogram)$.
			
			\begin{exa}
			\label{ex:fc-datalog-program}
			The $\fcdatalog$ program $\fcdatalogprogram = (\inputsymbol,\{\ans, R\}, \fcdatalogruleset)$ where $\fcdatalogruleset$ is:
			\begin{align*}
				\ans() &\shortleftarrow \inputsymbol \doteq yy\rc R(y); &
				R(x) &\shortleftarrow x \doteq y\mathsf{a}\rc R(y);\\		
				R(x) &\shortleftarrow x \doteq \emptyword;&
				R(x) &\shortleftarrow x \doteq y\mathsf{b}\rc R(y).
			\end{align*}
			defines the language $\lang(\fcdatalogprogram) \coloneqq \{ v \cdot v \mid v \in \{\mathsf{a},\mathsf{b}\}^\ast\}$. The $\fcdatalog$ program $\anotherfcdatalogprogram$ that has only the single rule $\ans() \shortleftarrow \inputsymbol \doteq yy \textrc R(y)$ defines the language $\lang(\anotherfcdatalogprogram) \coloneqq \{v \cdot v \mid v \in \terminals^{\ast}\}$.
			\end{exa}
			
			\begin{rem}
				As in \cite{FC}, it is not required for the universe variable $\inputsymbol$ to appear in a rule, as long as $\sub(\inputsymbol)$, and thus the universe, is defined. An example of such a program is defined in the proof of Lemma \ref{lem:linear-combined-pspace-hard}. A rule of thumb from \cite{FC} is that $\inputsymbol$ is only required when referring to some ``global'' property of the entire input word $w$. An example of such a property is expressing that the input word is a square, as in Example \ref{ex:fc-datalog-program}.
			\end{rem}
			
			We examine the \emph{model checking} problem for $\fcdatalog$, that is given a Boolean $\fcdatalog$ program $\fcdatalogprogram$ and a word $w$, decide if $w \in \lang(\fcdatalogprogram)$. We consider three perspectives: \emph{data complexity}, where the program $\fcdatalogprogram$ is fixed and only the word $w$ is considered  input, \emph{expression complexity}, where the word $w$ is fixed and only the program $\fcdatalogprogram$ is considered  input, and \emph{combined complexity}, where both the word $w$ and program $\fcdatalogprogram$ are considered  input (for details on these perspectives, see e.g. \cite{Libkin}). 
			
			We call a set of $\fcdatalog$ programs a \emph{fragment} (of $\fcdatalog$). We say that a fragment $\mathcal{F}$ \emph{captures} a complexity class $\mathbb{C}$ if $\mathbb{C} = \{\mathcal{L}(\fcdatalogprogram) \mid \fcdatalogprogram \in \mathcal{F}\}$. Note that, \changetwo{following directly from the definitions,} if $\mathcal{F}$ captures $\mathbb{C}$, then the data complexity of $\mathcal{F}$ \changetwo{is} $\mathbb{C}$. The only complexity result that is known for $\fcdatalog$ is that $\fcdatalog$ captures~$\p$ (see Theorem 4.11 of \cite{FC}), however, there are results for other versions of $\fc$, such as those discussed in Section \ref{sec:introduction}.
		
	\section{Efficient FC-Datalog}
		\label{sec:efficient}
		$\fcdatalog$ captures $\p$, and as $\p$ is often not considered efficient for data complexity, this presents an opportunity for optimization. We show that there is also such an opportunity for combined complexity.

		\begin{thm}
			\label{thm:exptime-complete}
			\changeother{C}ombined complexity of $\fcdatalog$ is $\exptime$-complete. 
		\end{thm}
		We show this with two intermediate results, Lemma \ref{lem:exptime-upper-bound} and Lemma \ref{lem:exptime-hard}. For the $\exptime$ upper bound, using the naive $\fcdatalog$ evaluation algorithm is enough. It is straightforward to see that as our universe is finite, the number of tuples we can add to the relations is exponential, and that the naive evaluation algorithm consists of a loop with an exponential bound on the number of iterations and an exponential number of steps in each iteration. 
		\begin{lem}
			\label{lem:exptime-upper-bound}
			\changeother{C}ombined complexity of $\fcdatalog$ is in $\exptime$.
		\end{lem}
		\begin{proof}
			We begin with all relations initialized to $\emptyset$. As the universe is finite, we can run a loop that, for every relation symbol $R$ and every tuple $\Vec{t}$ that is not in the relation $R$, attempts to add $\Vec{t}$ to $R$. We then terminate this loop when nothing can be added. The number of tuples $\Vec{t}$ is exponential in the size of the input. As every iteration of this loop adds at least one tuple, we have an exponential bound on the number of iterations of this loop. Also, as each iteration can add an exponential number of tuples, we have an exponential bound on the number of steps of each iteration. Therefore, evaluation is in $\exptime$.
		\end{proof}
		
		For $\exptime$-hardness, we use a reduction from classical $\datalog$ evaluation, which is again straightforward.

		\begin{lem}
			\label{lem:exptime-hard}
			\changeother{C}ombined complexity of $\fcdatalog$ is $\exptime$-hard.
		\end{lem}
		\begin{proof}
			Given a $\datalog$ program $\fcdatalogprogram'$ and an input database $\inputdatabase$, we construct an $\fcdatalog$ program $\fcdatalogprogram$ and a word $w$ such that $\fcdatalogprogram$ accepts $w$ if and only if $\fcdatalogprogram'$ accepts $\inputdatabase$.
			
			We can use factors of $w$ to represent elements of the active domain of $(\fcdatalogprogram', \inputdatabase)$, that is all the tuples in the relations in $\inputdatabase$ and all the constants in $\fcdatalogprogram'$. We can do this by encoding each element of the active domain as a unary factor of $w$. We can therefore replace constants in the rules of $\fcdatalogprogram'$ directly with their encoding, and for every tuple $\Vec{t}$ in every relation $R$ we add a new rule that adds the encoding of $\Vec{t}$ to the relation $R$.
		\end{proof}

		Our goal in this section is to identify fragments with lower data and combined complexity. In Section \ref{subsec:linearity} we employ linearity from classical $\datalog$, which improves data complexity to $\nlogspace$ and combined complexity to $\pspace$. Then in Section \ref{subsec:determinism} we define the deterministic linear fragment which improves data complexity to $\logspace$. However, checking membership of this fragment is expensive.
		
		\subsection{Linearity}
			\label{subsec:linearity}
			\changeother{There exists a fragment} of classical $\datalog$ that captures the complexity class $\nlogspace$ on ordered structures (see e.g. \changetwo{\cite{Gottlob2003}}), namely the fragment of all \changethree{semi-positive} \emph{linear} programs. As discussed in Section \ref{sec:introduction}, there exists a more general definition for linearity (see e.g \cite{Alice}) that does not affect the complexity. This restriction being that every rule is permitted at most one atom in the body that has an intensional relation symbol which is mutually recursive with the head relation symbol. We now translate this from the relational setting to the text setting.

			For an $\fcdatalog$ program $\fcdatalogprogram = (\inputsymbol, \relationsymbols, \fcdatalogruleset)$, we define a relation $R \shortleftarrow R'$ over relation symbols $R, R' \in \relationsymbols$ if there exists a rule $\fcdatalogrule \in \fcdatalogruleset$ where $R$ is the head relation symbol and $R'$ appears in the body. We then denote the transitive closure of $\shortleftarrow$ by $\overset{+}{\shortleftarrow}$, and two relation symbols $R$ and $R'$ are \emph{mutually recursive} if $R$ $\overset{+}{\shortleftarrow}$ $R'$ and $R'$ $\overset{+}{\shortleftarrow}$ $R$.
			
			\begin{defi}
				\label{def:linear-fc-datalog}
				\changetwofour{Let $\fcdatalogprogram$ be an $\fcdatalog$ program with rule set $\fcdatalogruleset$. A rule $\fcdatalogrule \in \fcdatalogruleset$ is \emph{linear} with respect to $\fcdatalogprogram$ if at most one atom in the body of $\fcdatalogrule$ has a relation symbol with which the head relation symbol of $\fcdatalogrule$ is mutually recursive. If every $\fcdatalogrule \in \fcdatalogruleset$ is linear, then $\fcdatalogprogram$ is \emph{linear}}.
			\end{defi}
			
			The two programs we have seen so far are both linear. Before analyzing the complexity of this fragment, we demonstrate an example of how linearity can be violated.

			\begin{exa}
				\label{ex:linear-fcdatalog}
				The $\fcdatalog$ programs given in Example \ref{ex:fc-datalog-program-intro} and Example \ref{ex:fc-datalog-program} are both linear $\fcdatalog$ programs. An example $\fcdatalog$ program that is not linear is the following program that retrieves all even length factors of the input word:
				\begin{align*}
					\ans(z) &\shortleftarrow z \doteq xy \rc \ans(x) \rc \ans(y); \\
					\ans(z) &\shortleftarrow z \doteq xy \rc L(x) \rc L(y);
				\end{align*}
				and a rule 	$L(x) \shortleftarrow x \doteq \mathsf{a}$ for each $\mathsf{a} \in \terminals$. This is not linear as $\ans$ is mutually recursive with itself, and the first rule has $\ans$ as its head relation symbol and two atoms in the body that have the relation symbol $\ans$.
			\end{exa}
			
			Checking if a given $\fcdatalog$ program is linear can be done in linear time with respect to the program's number of rules, as it amounts to determining the \emph{strongly connected components} of a directed graph. A \emph{strongly connected component} (SCC) of a graph is a maximal subset of nodes in which every node is reachable from every other node.

			\begin{prop}
				The linearity of an $\fcdatalog$ program can be decided in linear time.
			\end{prop}
			\begin{proof}
				Let $\fcdatalogprogram = (\inputsymbol, \relationsymbols, \fcdatalogruleset)$ be an $\fcdatalog$ program. Initialize $E = \emptyset$. We parse every rule of $\fcdatalogruleset$ and for all pairs $R_1, R_2 \in \relationsymbols$ where $R_1 \shortleftarrow R_2$, add $(R_1, R_2)$ to $E$. We can do this in $\bigo(\fcdatalogruleset)$ time. We then build the graph~$\mathcal{G}_\fcdatalogprogram = (\relationsymbols, E)$ and use Tarjan's strongly connected components algorithm (see Theorem 13 of \cite{Tarjan}), to compute the strongly connected components of $\mathcal{G}_\fcdatalogprogram$ in time $\bigo(\abs{\relationsymbols}+\abs{E})$. As in \cite{Tarjan}, each SCC has a root node. We can thus associate each element of a SCC with its root node in a dictionary.
				
				We then perform a pass over each rule in $\fcdatalogruleset$. If the number of relation atoms in the body is less than two, we proceed to the next rule. Otherwise, we look up in our dictionary the SCC roots of the head relation symbol and the relation symbol of every relation atom in the body. For every body relation symbol, we check if its SCC root node is the same as the SCC root node of the head relation symbol, and maintain a counter for the number of cases where it is. If this counter exceeds one, we reject. Otherwise, we proceed to the next rule. We can therefore achieve this in $\bigo(\fcdatalogruleset)$ time.
			\end{proof}

			Moreover, whilst unrestricted $\fcdatalog$ captures $\p$, the restriction to linear $\fcdatalog$ substantially improves the data complexity.
			
			\begin{thm}
				\label{thm:capture-nlogspace}
				Linear $\fcdatalog$ captures $\nlogspace$.
			\end{thm}
			
			We show the two directions of the proof in Lemma \ref{lem:membership-nlogspace} and Lemma \ref{lem:all-nlogspace}. We first obtain the $\nlogspace$ upper bound by building a nondeterministic algorithm with a constant number of pointers. See e.g. Lecture 5 of \cite{Kozen06} for this characterization of $\nlogspace$.

			\begin{lem}
				\label{lem:membership-nlogspace}
				\changeother{D}ata complexity of linear $\fcdatalog$ is in $\nlogspace$.
			\end{lem}
				
			\begin{proof}
				Let $w$ be the input word. Fix a linear $\fcdatalog$ program $\fcdatalogprogram = (\inputsymbol, \relationsymbols, \fcdatalogruleset)$. Our pointers are modelled as positions in $w$. We perform the procedure recursively by rule, as with the definition of $\fcdatalogprogram$. In each step, for each rule~$\fcdatalogrule \in \fcdatalogruleset$, we represent each free variable~$x \in \free(\fcdatalogrule)$ by two pointers, representing the start and end of an occurrence of $\sub(x)$. As $\fcdatalogprogram$ is fixed, we have a fixed number of free variables, and therefore a fixed number of pointers.
				
				A rule $\fcdatalogrule \in \fcdatalogruleset$ with head relation symbol $H$ is nondeterministically ``guessed'', as are the values of all free variables $x \in \free(\fcdatalogrule)$. In the first iteration we have $H = \ans$. Deciding if we accept can take up to three steps and at any point, if an atom doesn't hold then we reject.
					
				In the first step, we check all of the pattern equation atoms. For every atom $\atom \in \fcdatalogrule$ that is a pattern equation $x \doteq \alpha$, we need to check if $\sub(x) = \sub(\alpha)$. We do this by, inside the part of $w$ that represents $\sub(x)$, trying to match all positions $\alpha_{i} \in (\vars \cup \terminals)$ of the pattern $\alpha$. This requires an additional constant number of pointers, such as the position $i$ and the location inside $\sub(\alpha_{i})$. If all of these hold and there does not exist any relation atoms in the body of~$\fcdatalogrule$, then we accept. Otherwise, we proceed to the next step. 
				
				In the second step, we check all of the relation atoms of the form $N(\Vec{x})$ where $N$ is not mutually recursive with $H$. Each of these is solved as a subroutine where the set of relation symbols is $\relationsymbols' \subseteq \relationsymbols \setminus \{H\}$. If all of these hold and there does not exist an atom of the form $M(\Vec{x})$ where $M$ is mutually recursive with the head relation symbol, then we accept. Otherwise, we proceed to the next step. 
				
				In the third step, we check the relation atom of the form $M(\Vec{x})$. We nondeterministically ``guess'' a rule $\anotherfcdatalogrule$ with head relation symbol $M$, and retain only the values of the free variables of $\fcdatalogrule$~that are passed to $\anotherfcdatalogrule$. We then perform a further iteration of these three steps for $\anotherfcdatalogrule$, and where $H = M$.
					
				This process maps the steps of the proof tree, and if the word is accepted, then there exists some sequence of steps such that when the final step is reached, the program will terminate. We see that each time, we save only the free variables of the rule in question, and it is therefore possible to recurse an unbounded amount of times whilst maintaining a constant number of pointers.	
			\end{proof}

			\begin{rem}
				\label{rem:loop-detection}
				The astute reader may notice that the algorithm given in the proof of Lemma~\ref{lem:membership-nlogspace} may enter an infinite loop. However, it is well-known (see e.g. Theorem 4.2 of \cite{AroraBarak}) that if we have only $\log n$ space, then the algorithm can run for only $2^{\bigo(\log n)}$ time until it enters a repeated configuration and hence an infinite loop. Thus, as in Remark 4.3 of \cite{AroraBarak}, we can simply use a fixed number of pointers to maintain a step count, and if the step count exceeds this bound, we reject.
			\end{rem}

			We show the other direction, that we can express all of $\nlogspace$, by demonstrating how we can simulate multi-headed two-way nondeterministic finite automata, which are equivalent to nondeterministic Turing machines with logarithmic space (see e.g. \cite{Hartmanis72, Kozen06}). See e.g. \cite{Holzer} for a survey on the computational and descriptional complexity of two-way finite automata. We discuss the connection between automata and $\fcdatalog$ further in Section~\ref{subsec:comparison}.
			
			\begin{defi}
				\label{def:2nfak}
				A \emph{two-way nondeterministic finite automaton with k heads} ($2$NFA($k$)) is a structure $M=(Q, \terminals, \updelta, H, q_{0}, F)$ where:
				\begin{itemize}
					\item $Q$ is a finite set of states, $Q \neq \emptyset$,
					\item $\terminals$ is the input alphabet that does not contain the special symbols $\cent$ and $\mathdollar$,
					\item $\updelta$ is the transition function that maps $Q \times \terminals \cup \{\cent,\mathdollar\}$ into $\mathcal{P}(Q) \times \{-1,0,1\}$, with the restriction that for all $p,q \in Q$, if $(q,d) \in \updelta(p,\cent)$ this implies that $d \geq 0$, and if $(q,d) \in \updelta(p,\mathdollar)$ this implies that $d \leq 0$,
					\item $H$ is the head selector function, $H \colon Q \shortrightarrow \{1,2,\ldots,k\}$,
					\item $q_{0} \in Q$ is the initial state, and
					\item $F \in Q$ is the accepting state.
				\end{itemize}
				If $(q,d) \in \updelta(p,\mathsf{a})$ and $H(p) = h$, then if $M$ is in state $p$ with head $h$ scanning $\mathsf{a}$ on the input tape, $M$ may enter state $q$ and move $h$ to the right $d$ spaces. Several heads of $M$ may scan the same input tape space simultaneously, and the heads may move freely past one another. As in \cite{Holzer}, we can assume that the automaton has endmarkers, and that there is exactly one halting state in which we accept.
			\end{defi}

			As it is convenient to do so, we now define an explicit construction from a $2$NFA($k$) to a linear $\fcdatalog$ program.
			
			\begin{lem}
				\label{lem:all-nlogspace}
				Every $\nlogspace$ language can be expressed as a linear $\fcdatalog$ program.
			\end{lem}
			\begin{proof}
				Let $M$ be a 2NFA($k$) and let $w$ be the input word. We model the positions of the $k$ heads of $M$ as prefixes of $w$. We firstly handle the special case when $w=\emptyword$, where the heads cannot move. We simply evaluate if $\emptyword \in \lang(M)$, and if so include the rule $\ans() \shortleftarrow \inputsymbol \doteq \emptyword$. If $\emptyword \notin \lang(M)$, nothing extra is required. 
					
				We simulate the accepting path through the automaton with the top-down evaluation of the program. For every state~$q_i \in Q$ we define a $k$-ary relation $Q_{i}$ that expresses the possible head positions $M$ can have when in state $q$. We first construct an initialization rule for the start state $Q_0$ with all $k$ heads in the leftmost position: 
				\begin{equation*}
					\ans() \shortleftarrow Q_{0}(\emptyword,\ldots,\emptyword).
				\end{equation*}
				As this rule has only one relation atom in its body, it trivially has at most one relation atom in its body which has a relation symbol that is mutually recursive with the head relation symbol. Hence, it is linear. We then translate the transition function $\updelta$ of $M$ to a set of $\fcdatalog$ rules. When $M$ is in state $q_n$, we have $Q_{n}(x_{1},\ldots,x_{k})$, where $x_{1},\ldots,x_{k}$ are the prefixes indicating the positions of the $k$ heads. If $w = x_{1}\mathsf{a}_{1}z_{1},\ldots,w = x_{k}\mathsf{a}_{k}z_{k}$, where $\mathsf{a}_{1}, \ldots, \mathsf{a}_{k} \in \terminals$, then we read $\mathsf{a}_{1},\ldots,\mathsf{a}_{k}$, move to successor state $q_m$ and adjust the heads accordingly. We then have $Q_{m}(y_{1},\ldots,y_{k})$ where, for $1 \leq i \leq k$:
				\begin{itemize}
					\item if head $i$ moves right, denoted $(i, \shortrightarrow)$, then $y_{i}\doteq x_{i}\mathsf{a}_{i}$,
					\item if head $i$ moves left, denoted $(i, \shortleftarrow)$, then $x_{i}\doteq y_{i}\mathsf{a}_{i}$
					\item if head $i$ does not move, denoted $(i, -)$, then $y_{i}\doteq x_{i}$.
				\end{itemize}
				This translation is therefore mapped to the linear $\fcdatalog$ rule:
				\begin{multline*}
					Q_{n}(x_{1}, \ldots, x_{k}) \shortleftarrow Q_{m}(y_{1},\ldots,y_{k})\rc 
					\bigwedge_{j=1}^{k} \inputsymbol \doteq x_{j}\mathsf{a}_{j}z_{j} \rc \\
					\bigwedge_{\ell=1, (\ell, \shortrightarrow)}^{k} y_{\ell} \doteq x_{\ell}\mathsf{a}_{\ell} \rc
					\bigwedge_{m=1, (m, \shortleftarrow)}^{k} x_{m}\doteq y_{m}\mathsf{a}_{m}\rc 
					\bigwedge_{n=1, (n, -)}^{k} y_{n}\doteq x_{n}.
				\end{multline*}
				For $1 \leq j \leq k$, the value of $z_j$ is the suffix of $w$ that makes the equation $w \doteq x_j \mathsf{a}_j z_j$ true; this ensures that every $x_j$, and thus every $y_j$ is a prefix of $w$. For the same reason as the initialization rule, this rule is linear. We then repeat this for all transitions $q_n$ to $q_m$ where $q_m \neq F$. 
				
				For all transitions $q_n$ to $q_m$ where $q_m = F$, we map these to linear $\fcdatalog$ rules as for when $q_m \neq F$, except without the atom $Q_m(y_1, \ldots, y_k)$. Each of these rules have zero relation atoms in their bodies, and so are also linear. We see that if $w$ is accepted, then $\inputsymbol$ is in the relation representing the final state and we trace back the path through the automaton from the final state to the start state, where each time the prefix of $w$ is in the appropriate state relation.
			\end{proof}
			
			In addition to the results for data complexity, the restriction to linear $\fcdatalog$ also substantially improves combined complexity, which we recall is $\exptime$-complete for unrestricted $\fcdatalog$.
			
			\begin{thm}
				\label{thm:linear-combined-complexity}
					Combined and expression complexity of linear $\fcdatalog$ are $\pspace$-complete.
			\end{thm}

			We give both upper bounds in Lemma \ref{cor:linear-combined-membership-pspace} and obtain the lower bounds in Lemma~\ref{lem:linear-combined-pspace-hard} and Lemma~\ref{lem:linear-expression-pspace-hard} for combined and expression complexity  respectively. Membership of $\pspace$ can be established by giving a nondeterministic polynomial space algorithm, which by Savitch’s theorem that $\pspace = \npspace$ (see e.g. \cite{Alice}) can be transformed into a deterministic polynomial space algorithm. We now show the $\pspace$ upper bounds for combined and expression complexity. 

			\begin{lem}
				\label{cor:linear-combined-membership-pspace}
				\changeother{Combined and expression} complexity of linear $\fcdatalog$ is in $\pspace$.
			\end{lem}
			\begin{proof}
				This follows from the proof of Lemma \ref{lem:membership-nlogspace}, as if we allow a polynomial number of pointers then the proof is the same mutatis mutandis.
			\end{proof}

			We obtain $\pspace$-hardness for combined complexity by giving a polynomial-time reduction from the following well-known $\pspace$-complete problem: 
			
			\begin{defi}
				\label{def:pspace-problem}
				The problem ``given a deterministic Turing machine~$T$ and an integer $k$ encoded in unary, decide whether $T$ accepts $\emptyword$ in space $k$'' is $\pspace$-complete. As in~\cite{Gottlob2003}, we can assume w.l.o.g. that such a deterministic Turing machine accepts if and only if it halts in a special state $\Omega$ with a completely blank tape and the head is in the leftmost position.
			\end{defi}
			
			We now reduce this problem to combined complexity of linear $\fcdatalog$. 

			\begin{lem}
				\label{lem:linear-combined-pspace-hard}
				\changeother{C}ombined complexity of linear $\fcdatalog$ is $\pspace$-hard.
			\end{lem}
			\begin{proof}
				We show a polynomial time reduction from the problem given in Definition \ref{def:pspace-problem}. We define a polynomial time function that maps $T$ and $k$ into a Boolean linear $\fcdatalog$ program $\fcdatalogprogram$ and a word $w$, such that $w$ is accepted by $\fcdatalogprogram$ if and only if $T$~accepts $\emptyword$ without leaving the first $k$ tape cells.
				
				Assume $T$ has a tape alphabet $\Gamma$, where $\abs{\Gamma} = g$, and a state set $Q$, where $\abs{Q} = s$. We define $w = \mathsf{a}^{n}$, where $n = \max{(k,g)}$. We can therefore encode each tape symbol and head position as a prefix of $w$ by encoding every $i$, for $0 \leq i \leq n$, as $\mathsf{a}^{i}$. The tape position $i$ is represented by $\mathsf{a}^{i}$, as in \cite{Gottlob2003}, and each tape letter is encoded as an $\mathsf{a}^{i}$ that is unique for its type, the exact number is immaterial. We represent the encoding of an object $o$ by $[o]$. We represent a configuration of $T$ as an $\fcdatalog$ atom of the form $C_{q}(head, cell_{1},\ldots, cell_{k})$ where $q$ is the state of $T$, $head$ is a unary encoding of the head position of $T$, and for $0 \leq i \leq k$, $cell_{i}$ is a unary encoding of tape cell $i$.
				
				We now describe a linear $\fcdatalog$ program $\fcdatalogprogram$ that simulates the computation steps of $T$ when started with a blank tape (in other words, with an empty input). The program $\fcdatalogprogram$ consists of an initialization rule, a number of transition rules, and an accepting rule. We see that each rule only has one relation atom in its body and so $\fcdatalogprogram$ is linear.
				Let $q_{0}$ represent the initial state, $h_{0}$ represent the initial head position, and $\flat$ represent the blank symbol. The initialization rule is:
				\begin{equation*}
					\ans() \shortleftarrow C_{q_0}(h, c_{1},\ldots, c_{k}) \rc h \doteq [h_{0}]\rc \bigwedge_{j=1}^{k} c_{j} \doteq \flat.
				\end{equation*}
				
				For every transition $t$ of $T$, and each head position from which $t$ is possible, we define a corresponding rule in $\fcdatalogprogram$. The transition: ``\emph{if $T$ is in state $q_{\ell}$ and we read $\mathsf{c}$, then write $\mathsf{b}$, move the head right from position $i$ to position $i+1$, and enter state $q_{m}$}" is represented by the linear $\fcdatalog$ rule:
				\begin{multline*}
					C_{q_\ell}(h, c_{1}, \ldots, c_{k}) \shortleftarrow C_{q_m}(h', c'_{1},\dots,c'_{k})\rc h \doteq \mathsf{a}^{i}\rc \\
					h' \doteq \mathsf{a}^{i+1}\rc c_{i} \doteq [\mathsf{c}]\rc c'_{i} \doteq [\mathsf{b}]\rc  \bigwedge_{j=1, j\neq i}^{k} c_{j} \doteq c'_{j}.
				\end{multline*}
				If the head was moved left, then the rule would be the same apart from $h' \doteq \mathsf{a}^{i-1}$ instead of $h' \doteq \mathsf{a}^{i+1}$. 	If the head was moved left, then the rule would be the same apart from $h' \doteq \mathsf{a}^{i-1}$ instead of $h' \doteq \mathsf{a}^{i+1}$. Finally, the accepting rule accepts $w$ if the conditions for $T$ to halt on input $\emptyword$ in space $k$ are met. As in \cite{Gottlob2003}, w.l.o.g. these are: being in the special state $\Omega$, the head being in the leftmost position, and the tape being completely blank. The accepting rule is therefore:
				\begin{equation*}
					C_{\Omega}(h, c_{1},\dots,c_{k}) \shortleftarrow h \doteq \emptyword \rc \bigwedge_{j=1}^{k} c_{j} \doteq \flat.
				\end{equation*}
				
				We see that if the Turing machine accepts, then we trace back the appropriate configurations from the accepting state to the start state. Each rule is linear in the size of the unary encoding of $k$, and the number of possible rules is $kgs$, and thus polynomial in the size of $k$, $g$ and $s$. We can assume  $T$ has size $gs$, as in \cite{Alice}, and the length of $w$ is linear in the size of $k$ and $T$. Consequently, $\fcdatalogprogram$ can be constructed in polynomial time, and we thus have our polynomial time reduction.
			\end{proof}
			
			We now show $\pspace$-hardness for expression complexity by adapting the proof of Lemma \ref{lem:linear-combined-pspace-hard}.

			\begin{lem}
				\label{lem:linear-expression-pspace-hard}
				\changeother{E}xpression complexity of linear $\fcdatalog$ is $\pspace$-hard.
			\end{lem}
			\begin{proof}
				We modify the proof of Lemma \ref{lem:linear-combined-pspace-hard} by fixing the input word $w = \mathsf{a}$. We again show a polynomial time reduction from the problem given in Definition \ref{def:pspace-problem}. Let $T$ have a tape alphabet $\Gamma$, and by standard techniques we can assume w.l.o.g. that $\abs{\Gamma} = 2$. Also let $T$ have a state set $Q$, where $\abs{Q} = s$, and let $T$ have $v$ possible transitions. Since $w = \mathsf{a}$, we have two factors of $w$, namely $\emptyword$ and~$\mathsf{a}$.
					
				We can therefore encode each head position as a binary vector $head = h_1, \ldots, h_{\lceil \log k \rceil}$ and each tape symbol as a binary symbol. We again represent the encoding of an object $o$ by $[o]$. We represent a configuration of $T$ as an $\fcdatalog$ atom of the form $C_{q}(head, cell_{1},\ldots, cell_{k})$ where $q$ is the state of $T$, $head$ is a binary encoding of the head position of $T$, and for $0 \leq i \leq k$, $cell_{i}$ is a binary encoding of tape cell $i$.
					
				We now describe a linear $\fcdatalog$ program $\fcdatalogprogram$, simulating the evolution of $T$ when started with a blank tape. We again see that each rule only has one relation atom in its body and so $\fcdatalogprogram$ is linear. Let $q_{0}$ represent the initial state, $h_{1}, \ldots h_{k}$ represent the initial head position encoded in binary, and $\flat$ represent the blank symbol. The initialization rule is:
					
				\begin{equation*}
					\ans() \shortleftarrow C_{q_0}(h_1, \ldots, h_{\lceil \log k \rceil}, c_{1},\ldots, c_{k}) \rc \bigwedge_{j=1}^{k} c_{j} \doteq \flat.
				\end{equation*}
		
				For every transition $t$ of $T$, we define corresponding rules in $\fcdatalogprogram$. The transition: ``\emph{if $T$ is in state $q_{\ell}$ and we read $\mathsf{\flat}$, then write $\mathsf{b}$, move the head right from position $i$ to position $i+1$, and enter state $q_{m}$}" is represented by linear $\fcdatalog$ rules of the form:
				\begin{multline*}
					C_{q_\ell}(h_1, \ldots, h_{\lceil \log k \rceil}, c_{1}, \ldots, c_{k}) \shortleftarrow C_{q_m}(h'_1, \ldots, h'_{\lceil \log k \rceil}, c'_{1},\dots,c'_{k})\rc
					h_{n} \doteq 0 \rc \bigwedge_{i=n+1}^{\lceil \log k \rceil} h_{i} \doteq 1,\\
					h'_{n} \doteq 1 \rc \bigwedge_{j=1}^{n-1} h'_{j} \doteq h_{j} \rc  \bigwedge_{\ell = n+1}^{\lceil \log k \rceil} h'_{\ell} \doteq 0 \rc c_{m} \doteq \flat \rc c'_{m} \doteq [\mathsf{b}] \rc  \bigwedge_{p=1, p\neq m}^{k} c_{p} \doteq c'_{p}.
				\end{multline*}
				We have a copy of this rule for each possible value of $n$, for $1 \leq n \leq \lceil \log k \rceil$, and therefore have a rule that increments every possible head position. If the head moves left, then the rule performs a binary decrement rather than an increment. In this case we would have rules of the form:
				\begin{multline*}
					C_{q_\ell}(h_1, \ldots, h_{\lceil \log k \rceil}, c_{1}, \ldots, c_{k}) \shortleftarrow C_{q_m}(h'_1, \ldots, h'_{\lceil \log k \rceil}, c'_{1},\dots,c'_{k})\rc
					h_{n} \doteq 1 \rc \bigwedge_{i=n+1}^{\lceil \log k \rceil} h_{i} \doteq 0,\\
					h'_{n} \doteq 0 \rc \bigwedge_{j=1}^{n-1} h'_{j} \doteq h_{j} \rc  \bigwedge_{\ell = n+1}^{\lceil \log k \rceil} h'_{\ell} \doteq 1 \rc c_{m} \doteq \flat \rc c'_{m} \doteq [\mathsf{b}] \rc  \bigwedge_{p=1, p\neq m}^{k} c_{p} \doteq c'_{p}.
				\end{multline*}
				We again have a copy of this rule for each possible value of $n$, for $1 \leq n \leq \lceil \log k \rceil$, and therefore have a rule that decrements every possible head position. Again, as in \cite{Gottlob2003}, w.l.o.g., the conditions for $T$ to halt on input $\emptyword$ in space $k$ are: being in the special state $\Omega$, the head being in the leftmost position, and the tape being completely blank. The accepting rule is therefore:
				\begin{equation*}
					C_{\Omega}(h_1, \ldots, h_{\lceil \log k \rceil}, c_{1},\dots,c_{k}) \shortleftarrow \bigwedge_{i=1}^{\lceil \log k \rceil} h_{i} \doteq 0 \rc \bigwedge_{j=1}^{k} c_{j} \doteq \flat.
				\end{equation*}
		
				Each rule in $\fcdatalogprogram$ has size is $\bigo(k \lceil \log k \rceil)$, and the number of possible rules is $\bigo(v\lceil \log k \rceil)$. We can assume $T$ has size $2s$, and the length of $w$ is fixed. We therefore see that $\fcdatalogprogram$ can be defined in polynomial time, and thus we have our polynomial time reduction.
			\end{proof}

			Therefore, through imposing linearity we have improved both data and combined complexity. On the other hand, we would like lower data complexity than $\nlogspace$, and lower combined complexity than $\pspace$. Furthermore, $\pspace$-completeness occurs even on a single-character input. As such, we look for more efficient fragments.
			
		\subsection{Determinism}
			\label{subsec:determinism}
			As linear $\fcdatalog$ captures $\nlogspace$, it is natural to look at causes of nondeterminism and see if there is a corresponding fragment that captures $\logspace$. For model checking, we evaluate $\fcdatalog$ programs \emph{top-down}.

			\begin{exa}
				\label{ex:top-down-eval}
				We show a top-down evaluation of the program $\fcdatalogprogram$ in Example \ref{ex:fc-datalog-program} on the input word $w = \mathsf{abab}$. The only rule we can apply first is $\ans() \shortleftarrow \inputsymbol \doteq yy \rc R(y)$. As $\sub(\inputsymbol) = w$, we see that $\sub(y) = \mathsf{ab}$, and we pass this into the relation $R$. We can then only apply the rule $R(x) \shortleftarrow x \doteq y\mathsf{b} \rc R(y)$. As $\sub(x) = \mathsf{ab}$, we have $\sub(y) = \mathsf{a}$ and we recurse on this value of $y$. We then apply $R(x) \shortleftarrow x \doteq y\mathsf{a} \rc R(y)$ and recurse on $\sub(y) = \emptyword$. We can then apply $R(x) \shortleftarrow x \doteq \emptyword$, and as this holds, we accept.
			\end{exa}
			
			Based on this top-down evaluation model, we define two categories of variables.

			\begin{defi}
				\label{def:input-output-vars}
				Let $\fcdatalogrule = R_h(x_1, \ldots x_m) \shortleftarrow R_m(y_1, \ldots y_n), \atom_1, \ldots, \atom_\ell$ be a linear $\fcdatalog$ rule where $R_m$ is mutually recursive with $R_h$, and for $1 \leq i \leq \ell$, the atom $\atom_{i}$ is either a pattern equation, or a relation atom with a relation symbol that is not mutually recursive with~$R_h$. Then the \emph{top variables} of $\fcdatalogrule$, denoted $\topvar(\fcdatalogrule) = (x_1, \ldots, x_m, \inputsymbol)$ and the \emph{bottom variables} of $\fcdatalogrule$, denoted $\bottomvar(\fcdatalogrule)=(y_1,\ldots,y_n)$.
			\end{defi}
			
			\changethree{In top-down evaluation, a rule's top variables contain values that were passed down from a preceding rule's evaluation, and its bottom variables contain the values that are passed down.} A variable can be both a top and bottom variable.

			\begin{exa}
				Let $\fcdatalogrule$ be the linear $\fcdatalog$ rule $R(x) \shortleftarrow x \doteq y\mathsf{a} \smallrc R(y)$ from Example~\ref{ex:fc-datalog-program}. Then $\topvar(\fcdatalogrule) = (x, \inputsymbol)$ and $\bottomvar(\fcdatalogrule) = (y)$.
			\end{exa}
			
			When evaluating a linear program, nondeterminism can occur in two ways: choosing the values of the variables when processing a rule, which we call \emph{local nondeterminism}, and choosing which rule to process, which we call \emph{global nondeterminism}. In order to eliminate all nondeterminism, we must ensure every program has both \emph{local} and \emph{global determinism}. To do this, we define the relation $W_{\fcdatalogrule}$ that holds the possible values for a rules top and bottom variables.

			\begin{defi}
				Let $w$ be the input word and let $\fcdatalogrule$ be a linear $\fcdatalog$ rule. Let the relation~$W_{\fcdatalogrule}$ be all pairs $(\sub(\topvar(\fcdatalogrule)), \sub(\bottomvar(\fcdatalogrule)))$ for all substitutions $\sub$ such that $(w, \sub)~\models~\we(\fcdatalogrule)$. For some $(\Vec{t_t}, \Vec{t_b}) \in W_{\fcdatalogrule}$, we call $\Vec{t_t}$ a \emph{top tuple} and $\Vec{t_b}$ a \emph{bottom tuple}.
			\end{defi}
			
			Recall that as that we are working under finite model semantics, each value is a factor of the word $w$. We now define our two criteria for \changeother{determinism.}

			\begin{defi}
				\label{def:determinism-criteria}
				A linear $\fcdatalog$ program $\fcdatalogprogram = (\inputsymbol, \relationsymbols, \fcdatalogruleset)$ is
				\begin{itemize}
					\item \emph{locally deterministic} if for each $\fcdatalogrule \in \fcdatalogruleset$, the relation $W_{\fcdatalogrule}$ is a partial function;
					\item \emph{globally deterministic} if for every relation symbol $R \in \relationsymbols$, for all pairs of distinct rules~${\fcdatalogrule, \anotherfcdatalogrule \in \fcdatalogruleset_{R}}$, there is no top tuple $\Vec{t}$ such that $(\Vec{t},\Vec{t_1}) \in W_{\fcdatalogrule}$ and $(\Vec{t},\Vec{t_2}) \in W_{\anotherfcdatalogrule}$, for~some bottom tuples $t_1$ and $t_2$.
				\end{itemize}
			\end{defi} 

			If $W_\fcdatalogrule$ is a partial function for some rule $\fcdatalogrule$, then top-down evaluation of $\fcdatalogrule$ is locally deterministic as for every top tuple there is at most one bottom tuple. If for all head relation symbols $R$, every top tuple is a valid input for at most one rule $\fcdatalogrule \in \fcdatalogruleset_{R}$, then we have global determinism as there is at most one rule that we can process.

			\begin{defi}
			\emph{Deterministic~linear~$\fcdatalog$} is the set of linear $\fcdatalog$ programs that are  locally and globally deterministic.
			\end{defi}

			Our condition \changetwo{is sufficiently strong as we precisely capture $\logspace$.}

			\begin{thm}
			\label{thm:deterministic-linear-capture-logspace}
			Deterministic linear $\fcdatalog$ captures $\logspace$.
			\end{thm}

			We show this with two intermediate results, Lemma \ref{lem:deterministic-in-logspace} and Lemma \ref{cor:deterministic-linear-all-logspace}. We first show a $\logspace$ upper bound by building a deterministic algorithm with a constant number of pointers. As with the nondeterministic case, see e.g. Lecture 5 of \cite{Kozen06} for this characterization.
			
			\begin{lem}
			    \label{lem:deterministic-in-logspace}
			   \changeother{D}ata complexity of deterministic linear $\fcdatalog$ is in $\logspace$.
			\end{lem}
			\begin{proof}
				Let $w$ be the input word. Fix a linear $\fcdatalog$ program $\fcdatalogprogram = (\inputsymbol, \relationsymbols, \fcdatalogruleset)$. As in Lemma \ref{lem:membership-nlogspace} our pointers are modelled as positions in $w$. In each step of the procedure, for each rule $\fcdatalogrule$ in $\fcdatalogruleset$, each free variable $x \in \free(\fcdatalogrule)$ is represented by two pointers, representing a start and end of an occurrence of $\sub(x)$. As $\fcdatalogprogram$ is fixed, we have a fixed number of variables and therefore a fixed number of pointers.
			
				We first take all the rules with head relation symbol $H$, where in the first iteration $H = \ans$. As $\fcdatalogprogram$ is globally deterministic, then there is at most one rule $\fcdatalogrule$ that holds. If there is no such rule, then we reject. As $\fcdatalogprogram$ is locally deterministic, then the input to $\fcdatalogrule$ is mapped to at most one output. We iterate through each possible value for each free variable $x \in \free(\fcdatalogrule)$ until we find either an accepting assignment or that no such assignment exists, and we do this in up to three steps. At any point, if an atom doesn't hold then we reject. 
				
				In the first step, we check all of the pattern equation atoms, as in the proof of Lemma~\ref{lem:membership-nlogspace}. If all of these hold and there does not exist any relation atoms in the body of $\fcdatalogrule$, then we accept. Otherwise, we proceed to the next step. 
				
				In the second step, we check all of the relation atoms of the form $\atom = N(\Vec{x})$ where $N$ is not mutually recursive with $H$. As in the proof of Lemma~\ref{lem:membership-nlogspace}, each of these are solved as a subroutine where the set of relation symbols is $\relationsymbols' \subseteq \relationsymbols \setminus \{H\}$. If all of these hold and there does not exist an atom of the form $M(\Vec{x})$ where $M$ is mutually recursive with the head relation symbol, then we accept. Otherwise, we proceed to the next step. 
				
				In the third step, we check the relation atom of the form $M(\Vec{x})$. We select a rule with head relation symbol $M$ that accepts the input $\Vec{x}$. As $\fcdatalogprogram$ is globally deterministic, then there is either no such rule, in which case we reject, or there is exactly one rule $\anotherfcdatalogrule$. We  then retain only the values of the free variables of $\fcdatalogrule$ that are passed to $\anotherfcdatalogrule$, and perform another iteration of this process for $\anotherfcdatalogrule$, and where $H = M$.
					
				As in the proof of Lemma \ref{lem:membership-nlogspace}, this process maps the steps of the proof tree, and if the word is accepted, then there is some sequence of steps such that when the final step is reached, the program will terminate. We see that each time, we save only the free variables of the rule in question and it is therefore possible to recurse an unbounded amount of times whilst maintaining a constant number of pointers.
			\end{proof}

			We can ensure the algorithm given in the proof of Lemma \ref{lem:deterministic-in-logspace} always terminates using the technique described in Remark \ref{rem:loop-detection} for Lemma \ref{lem:membership-nlogspace}. Furthermore, akin to the nondeterministic case, multi-headed two-way deterministic finite automata are equivalent to deterministic Turing machines with logarithmic space (see e.g. \cite{Hartmanis72, Kozen06}).

			\begin{defi}
				A \emph{two-way deterministic finite automaton with k heads} ($2$DFA($k$)) is a tuple $M=(Q, \terminals, \updelta, H, q_{0}, F)$ where:
				\begin{itemize}
					\item $Q$ is a finite set of states, $Q \neq \emptyset$,
					\item $\terminals$ is the input alphabet that does not contain the special symbols $\cent$  and $\mathdollar$,
					\item $\updelta$ is the transition function that maps $Q \times \terminals \cup \{\cent,\mathdollar\}$ into $Q \times \{-1,0,1\}$, with the restriction that for all $p,q \in Q$, if $(q,d) \in \updelta(p,\cent)$ this implies that $d \geq 0$, and if $(q,d) \in \updelta(p,\mathdollar)$ this implies that $d \leq 0$,
					\item $H$ is the head selector function, $H \colon Q \shortrightarrow \{1,2,\ldots,k\}$,
					\item $q_{0} \in Q$ is the initial state, and
					\item $F \in Q$ is the set of accepting states.
				\end{itemize}
				If $(q,d) \in \updelta(p,\mathsf{a})$ and $H(p) = h$, then if $M$ is in state $p$ with head $h$ scanning $\mathsf{a}$ on the input tape, $M$ may enter state $q$ and move $h$ to the right $d$ spaces. Several heads of $M$ may scan the same input tape space simultaneously, and the heads may move freely past one another. As in \cite{Holzer}, we can assume that the automaton has endmarkers, and that there is exactly one halting state in which we accept.
			\end{defi}

			We now give an explicit construction from a $2$DFA($k$) to a deterministic linear $\fcdatalog$ program. In fact, we can use the same construction as in the proof of Lemma~\ref{lem:all-nlogspace}.

			\begin{lem}
				\label{cor:deterministic-linear-all-logspace}
				Every $\logspace$ language can be expressed as a deterministic linear $\fcdatalog$ program.
			\end{lem}
			\begin{proof}
				Let $M$ be a $2$DFA($k$), and, as every $2$DFA($k$) is also a $2$NFA($k$), let $P$ be the linear $\fcdatalog$ program that simulates $M$, as defined using the construction in the proof of Lemma \ref{lem:all-nlogspace}. As we construct one rule for each transition, then each rule of $P$ is locally deterministic; each rule of $P$ maps one input to one output. Furthermore, as $M$ is a $2$DFA($k$), which is by definition deterministic, then $P$ is globally deterministic; there will be no input accepted by more than one of these rules.
			\end{proof}

			Unfortunately, where linearity for $\fcdatalog$ can be checked in polynomial time, checking determinism for linear $\fcdatalog$ is considerably more expensive, as the criteria we have to check are semantic rather than syntactic. In fact, checking nondeterminism of linear $\fcdatalog$ is as hard as the satisfiability problem for word equations. This problem has a lower bound of $\np$-hardness, which follows from the $\np$-hardness of the membership problem for pattern languages (see e.g. \cite{AngluinPatterns, Koscielski}), and an upper bound of nondeterministic linear space~(see \cite{Jez}); closing this gap is a longstanding open problem. We say a problem~$\probl$ is \emph{word equations-hard} if there exists a polynomial time reduction from an instance of the satisfiability problem for word equations to $\probl$. That is, there exists a polynomial function that transforms any word equation $E$ into an instance of $\probl$ such that $E$ is satisfiable if and only if $\probl$ accepts. Thus if $\probl$ could be solved in polynomial time, then so could word equation satisfiability.

			\begin{prop}    
				\label{lem:word-equations-hard}
				Checking local \changeother{or} global nondeterminism of linear $\fcdatalog$ is word equations-hard.
			\end{prop}
			
			\begin{proof}
				We first look at verifying a program is globally nondeterministic. Let $\alpha \doteq \beta$ be a word equation. Then let $P$ be the three rule $\fcdatalog$ program that consists of the rule $\ans() \shortleftarrow R_1(\inputsymbol)$ and the rules $\fcdatalogrule \coloneqq R_1(x) \shortleftarrow R_2(x)$ and $\anotherfcdatalogrule~\coloneqq~R_1(x) \shortleftarrow R_2(x) \rc y \doteq \alpha \rc y \doteq \beta$, for some $y \in \vars$. If $\alpha \doteq \beta$ is satisfiable, then there exists some substitution $\sub$ such that $(\sub(x), ()) \in W_\fcdatalogrule$ and $ (\sub(x), ()) \in W_\anotherfcdatalogrule$. Therefore $P$ is globally nondeterministic. On the other hand, if $\alpha \doteq \beta$ is not satisfiable, then for all substitutions $\sub$ we have $(\sub(x), ()) \in W_\fcdatalogrule$ and $(\sub(x), \vec{t_b}) \notin W_\anotherfcdatalogrule$, for any bottom tuple $t_b$. Therefore $P$ is globally deterministic.  
				
				We now look at verifying a program is locally nondeterministic. Let $\alpha \doteq \beta$ be a word equation. We build the conjunction $C \coloneqq (y \doteq \alpha) \wedge (y \doteq \beta) \wedge (z \doteq x\mathsf{a})$, for some $x, y, z \in \vars$ and some $\mathsf{a} \in \terminals$. Using the satisfiability-preserving reduction for conjunctions in e.g. Theorem~6 of~\cite{wordEquationExpPower}, we can reduce $C$ to a single word equation $E_C$. We then build the disjunction $D \coloneqq (z \doteq x) \vee E_C$. Using the satisfiability-preserving reduction for disjunctions in e.g. Theorem~6 of~\cite{wordEquationExpPower}, we can reduce $D$ to a single word equation~$E_D$ that contains two additional new variables $z', z'' \in \vars$. Thus, if $E_D$ is satisfiable, then $(z \doteq x) \vee (y \doteq \alpha \wedge y \doteq \beta \wedge z \doteq x\mathsf{a})$ is also satisfiable. As our alphabet $\terminals$ is fixed, this conversion is polynomial.
				
				Let $E_D = \alpha' \doteq \beta'$. We then build the $\fcdatalog$ program $\fcdatalogprogram$ that consists of the rule $\ans() \shortleftarrow R_1(\inputsymbol)$ and the rules $\fcdatalogrule \coloneqq R_1(x) \shortleftarrow R_2(z) \rc y' \doteq \alpha' \rc y' \doteq \beta'$ and $\anotherfcdatalogrule \coloneqq R_2(x) \shortleftarrow R_1(x)$. Then if $\alpha \doteq \beta$ is satisfiable, then there exists some substitution $\sub$ such that we have $(\sub(x), \sub(x)) \in W_{\fcdatalogrule}$ and $(\sub(x), \sub(x\mathsf{a})) \in W_{\fcdatalogrule}$, and hence $W_{\fcdatalogrule}$ is not a partial function. Therefore $\fcdatalogprogram$ is locally nondeterministic. On the other hand, if $\alpha \doteq \beta$ is not satisfiable, then for all substitutions~$\sub$, only $(\sub(x), \sub(x)) \in W_{\fcdatalogrule} $, and hence $W_{\fcdatalogrule}$ is a partial function. Therefore, $\fcdatalogprogram$ is locally deterministic. 
			\end{proof}
			
			Thus, although we have reduced the data complexity, we have lost the efficient checking of membership in the fragment. Furthermore, the determinism restriction does not improve the combined complexity.
			
			\begin{cor}
				Combined complexity of deterministic linear $\fcdatalog$ is $\pspace$-complete.
			\end{cor}
			\begin{proof}
				The $\pspace$ upper bound follows directly from Lemma \ref{cor:linear-combined-membership-pspace}. For $\pspace$-hardness, let $\fcdatalogprogram$ be the program defined using the construction in the proof of Lemma \ref{lem:linear-combined-pspace-hard}. Observe that $\fcdatalogprogram$ is deterministic.
			\end{proof}
 
			Finally, we show that in contrast to regular spanners, under standard complexity theoretic assumptions we cannot expect to have constant delay enumeration algorithms in our setting. A constant delay enumeration algorithm has a preprocessing phase, which often runs in linear time, and an enumeration phase that outputs solutions one by one, with constant time between any consecutive outputted solutions (see e.g. \cite{constantDelaySurvey}).

			\begin{prop}
				If there exists a constant delay enumeration algorithm \changeother{with polynomial preprocessing} for deterministic linear $\fcdatalog$, then $\p = \np$.
			\end{prop}
			\begin{proof}
				\changetwo{A constant delay enumeration algorithm with polynomial preprocessing for deterministic linear $\fcdatalog$ would, in polynomial time, return the first solution or terminate if there are no solutions. We can encode a pattern $\alpha$ and a word $w$ as a deterministic linear $\fcdatalog$ program $\ans() \shortleftarrow \inputsymbol \doteq \alpha$ where the input word is $w$. This means the existence of such an enumeration algorithm for deterministic linear $\fcdatalog$ would imply that we could decide the membership problem for pattern languages in polynomial time. As this problem is $\np$-hard (see \cite{AngluinPatterns}), such an enumeration algorithm cannot exist, unless $\p = \np$.}
			\end{proof}

			We therefore do not look further into constant delay enumeration. Instead we turn our focus to identifying a fragment that has efficient membership checking in addition to efficient data complexity, without compromising on expressive power.
		
	\section{A Framework for Efficient FC-Datalog}
		\label{sec:framework}
		In Section \ref{sec:efficient} we defined deterministic linear $\fcdatalog$, which allows us to reduce data complexity from $\exptime$ to $\logspace$ However, checking if a linear program is deterministic is as hard as the satisfiability problem for word equations. Therefore our main goal in this section is to define a further fragment that has efficient membership checking as well as efficient data complexity. We achieve this in Section \ref{subsec:olla}; the one letter lookahead restriction allows us to obtain $\logspace$ data complexity, and this time membership in the fragment can be checked in polynomial time. 
		
		On the other hand, combined complexity for this fragment remains $\pspace$-complete. In Section \ref{subsec:sd} we define a final restriction named strictly decreasing, which allows us to obtain linear combined complexity. We thus define the endpoints of an infinite range of $\fcdatalog$ fragments that all capture $\logspace$, and add the further dimension of strictly decreasing for linear combined complexity. As a result, fragments from this range can be tailored for particular applications, as we demonstrate in Section \ref{sec:application}. 
		
		Finally, in Section \ref{subsec:comparison}, we explain how our fragments of $\fcdatalog$, particularly those from our infinite range, can be viewed as generalizations of automata. 
		
		\subsection{One Letter Lookahead}
			\label{subsec:olla}
			In this subsection, we introduce a second fragment that captures $\logspace$, but where we can check membership in the fragment in polynomial time. This fragment has a severely restricted syntax but can still simulate multi-headed two-way deterministic finite automata, demonstrating how little of $\fcdatalog$ is required to retain this level of expressivity.
			
			\begin{defi}   
				\label{def:olla}
				Let $\fcdatalogrule$ be a linear $\fcdatalog$ rule. Let $x, \in \topvar(\fcdatalogrule)$, $y \in \bottomvar(\fcdatalogrule)$, and $\mathsf{a} \in \terminals \cup \{\emptyword\}$. The rule $\fcdatalogrule$ is a \emph{one letter lookahead} \text{($\olla$)} $\fcdatalog$ rule if each pattern equation has either of the below forms:
				\begin{enumerate}
					\item \label{case:emptyword-olla}$x \doteq \emptyword$ or $x \doteq x'$ where $x' \in \topvar(\fcdatalogrule)$.
					\item \label{case:xy}$x \doteq y\mathsf{a}$ or $x \doteq \mathsf{a}y$ (deleting a letter from a top variable).
					\item \label{case:yx}$y \doteq x\mathsf{a}$ or $y \doteq \mathsf{a}x$ (adding a letter to a top variable).
					\item for $z \in \vars$ where $z \notin \topvar(\fcdatalogrule) \cup \bottomvar(\fcdatalogrule)$ and $z$ does not occur elsewhere in $\fcdatalogrule$:
					\begin{enumerate}
						\item \label{case:uxaz}$\inputsymbol \doteq x\mathsf{a}z$ (matching the next letter in the input word $w$ after a top variable that represents a prefix of $w$).
						\item \label{case:uzax}$\inputsymbol \doteq z\mathsf{a}x$ (matching the previous letter in the input word $w$ before a top variable that represents a suffix of $w$).
					\end{enumerate}
				\end{enumerate}
			\end{defi}

			We now consider another way of categorising our permitted forms of pattern equations. Let $\fcdatalogrule$ be an $\olla$ $\fcdatalog$ rule. Intuitively, if a pattern equation $\atom \in \we(\fcdatalogrule)$ does not contain a terminal symbol $\mathsf{a} \in \terminals$, we call it \emph{$\emptyword$-$\olla$}. Otherwise, if some $\mathsf{a} \in \terminals$ occurs to the right of some $v \in \topvar(\fcdatalogrule) \cup \bottomvar(\fcdatalogrule)$ in $\atom$, then we call it \emph{right-$\olla$}, and if some $\mathsf{a} \in \terminals$ occurs to the left of some $v \in \topvar(\fcdatalogrule) \cup \bottomvar(\fcdatalogrule)$ in $\atom$, then we call it \emph{left-$\olla$}

			\begin{defi}
				\label{def:l-r-olla}
				Let $\atom$ be a pattern equation in an $\olla$ $\fcdatalog$ program. If the form of $\atom$ is that of case \eqref{case:emptyword-olla}, then we say $\atom$ is \emph{$\emptyword$-$\olla$}. If it is that of case \eqref{case:uxaz}, then we say $\atom$ is \emph{right $\olla$}. If it is that of case \eqref{case:uzax}, then we say $\atom$ is \emph{left $\olla$}. If the form of $\atom$ is that of case \eqref{case:xy}, then we say $\atom$ is \emph{right $\olla$} if $\atom = x \doteq y\mathsf{a}$ and $\atom$ is \emph{left $\olla$} if $\atom = x \doteq \mathsf{a}y$. Likewise, if the form of $\atom$ is that of case \eqref{case:yx}, then we say $\atom$ is \emph{right $\olla$} if $\atom = y \doteq x\mathsf{a}$ and $\atom$ is \emph{left $\olla$} if $\atom = y \doteq \mathsf{a}x$. 
			\end{defi}

			Definitions \ref{def:olla} and \ref{def:l-r-olla} give us everything that we need to define the one letter lookahead~($\olla$) fragment of $\fcdatalog$.

			\begin{defi}
				Let $\fcdatalogprogram = (\inputsymbol, \relationsymbols, \fcdatalogruleset)$ be a linear $\fcdatalog$ program. We say $\fcdatalogprogram$ is a \emph{one~letter~lookahead} \text{($\olla$)} $\fcdatalog$ program if:
				\begin{itemize}
					\item every $\fcdatalogrule \in \fcdatalogruleset$ is an $\olla$ $\fcdatalog$ rule; and
					\item for every relation symbol $R \in \relationsymbols$, we have that for every rule $\fcdatalogrule \in \fcdatalogruleset_{R}$, every variable ${v \in \topvar(\fcdatalogrule) \cup \bottomvar(\fcdatalogrule)}$ does not occur in both left $\olla$ and right $\olla$ pattern equations.
				\end{itemize}
			\end{defi}
			
			We can now use the notions of local and global determinism from Definition \ref{def:determinism-criteria} to obtain the deterministic $\olla$ ($\dolla$) $\fcdatalog$ fragment. 

			\begin{defi}
			An $\olla$ $\fcdatalog$ program is \emph{deterministic} (\emph{a $\dolla$ $\fcdatalog$ program}) if it is both locally and globally deterministic. 
			\end{defi}

			To ensure global determinism in $\olla$ $\fcdatalog$, we must not have pattern equations of the form $x \doteq x'$, for some $x, x' \in \topvar(\fcdatalogrule)$, that occur without a \emph{guard}.

			\begin{defi}
				Let $\fcdatalogprogram$ be an $\olla$ $\fcdatalog$ program. We say $\fcdatalogprogram$ is \emph{guarded} if for every relation symbol $R \in \relationsymbols$, we have that for any rule $\fcdatalogrule \in \fcdatalogruleset_{R}$, a pattern equation of the form $x \doteq x'$, for $x, x' \in \topvar(\fcdatalogrule)$, is in $\we(\fcdatalogrule)$ only in one of the following cases:
				\begin{itemize}
					\item if $\abs{\fcdatalogruleset_{R}} = 1$; or
					\item if another equation that has either of the forms $x=\emptyword$ or $x=\mathsf{a}y$ or $x=y\mathsf{a}$, for $y \in \bottomvar(\fcdatalogrule)$, is also in $\we(\fcdatalogrule)$.
				\end{itemize}
			\end{defi}

			Then by definition, every $\dolla$ $\fcdatalog$ program is guarded. In $\dolla$ $\fcdatalog$, we are no longer permitted to use pattern equations that perform an operation other than processing a string by at most one letter, or setting a string to $\emptyword$. For example, we can no longer use $x \doteq yy$ which splits a string in half.

			\begin{exa}
				\label{ex:not-dolla}
				The linear $\fcdatalog$ program $\fcdatalogprogram$ that is given in Example \ref{ex:fc-datalog-program} is not a $\dolla$~$\fcdatalog$ program as the rule $\ans() \shortleftarrow \inputsymbol \doteq yy \smallrc R(y)$ contains $ \inputsymbol \doteq yy$, which does not fit any of the permitted forms of pattern equations for an $\olla$ $\fcdatalog$ program.
			\end{exa}
			
			The simplest $\dolla$ $\fcdatalog$ program the authors could find that models the language of squares given in Example \ref{ex:fc-datalog-program} is the one that models the multi-headed two-way DFA; this program first identifies the middle of the string and processes the two halves letter by letter. Most, the authors included, would consider this to be impractical at best, which gives us our motivation for Section \ref{sec:application}; we consider $\dolla$ $\fcdatalog$ the most restricted endpoint of a range of fragments, and investigate further syntactical features that make writing programs more convenient without sacrificing desirable properties. Firstly though, with regard to the desirable properties of $\dolla$ $\fcdatalog$, we now see that despite restricting the programs substantially, $\dolla$ $\fcdatalog$ retains the expressive power of deterministic linear $\fcdatalog$.

			\begin{thm} 
				\label{thm:dolla-capture-logspace}
				$\dolla$ $\fcdatalog$ captures $\logspace$.
			\end{thm}
			The $\logspace$ upper bound follows directly from Lemma \ref{lem:deterministic-in-logspace}. We show that this fragment expresses all of $\logspace$ in Lemma \ref{lem:dolla-all-logspace}. Once again, we use our construction from the proof of Lemma~\ref{lem:all-nlogspace}, as the restriction permits just enough for a $\dolla$ $\fcdatalog$ program to simulate a $2$DFA($k$).
			
			\begin{lem}
				\label{lem:dolla-all-logspace}
				Every $\logspace$ language can be expressed as a $\dolla$ $\fcdatalog$ program.
			\end{lem}
			\begin{proof}
				Let $M$ be a $2$DFA($k$), and let $P$ be the deterministic linear $\fcdatalog$ program that simulates $M$, as defined in the proof of Lemma \ref{cor:deterministic-linear-all-logspace} using the construction in the proof of Lemma \ref{lem:all-nlogspace}. Observe that $P$ is an $\olla$ $\fcdatalog$ program. As $P$ is also deterministic, it is therefore a $\dolla$ $\fcdatalog$ program.
			\end{proof}

			Checking if a program is an $\olla$ $\fcdatalog$ program is straightforward. Checking  if an $\olla$ $\fcdatalog$ program is deterministic is easier than in the general case as the pattern equations are restricted to only match one letter at a time. For both local and global determinism, we need only consider the case where the rules do not contain contradicting pattern equations. If we have a rule with two pattern equations $\atom_1$ and $\atom_2$ that contradict each other, then as there are no solutions to $\atom_1 \wedge \atom_2$, this rule can never be applied.
			
			To check if a program is locally deterministic, we check that for every rule $\fcdatalogrule$ without pattern equations that contradict each other, every $x \in \topvar(\fcdatalogrule)$ appears in some pattern equation in $\we(\fcdatalogrule)$ or $x \in \bottomvar(\fcdatalogrule)$, and every $y \in \bottomvar(\fcdatalogrule)$ appears in some pattern equation in $\we(\fcdatalogrule)$ or $y \in \topvar(\fcdatalogrule)$. If so, then $W_{\fcdatalogrule}$ is a partial function. To check if a program is globally deterministic, we use \emph{profiles} for its rules.

			\begin{defi}
				Let $\fcdatalogrule$ be an $\olla$ $\fcdatalog$ rule that does not contain contradicting pattern equations. We define a \emph{profile} for $\fcdatalogrule$ as a function $pro_{\fcdatalogrule} \colon  \topvar(\fcdatalogrule) \shortrightarrow \terminals \cup \{  \changeother{\emptyword, \bot} \}$ where:
				\begin{equation*}
					pro_{\fcdatalogrule}(x) = \begin{cases}
						\mathsf{a} \text{ if there exists a pattern equation $x \doteq y\mathsf{a}$ or $x \doteq \mathsf{a}y$} \text{, for $\mathsf{a} \in \terminals$},\\
						\emptyword \text{ if there exists a pattern equation $x \doteq \changeother{\emptyword}$,}\\
						\bot \text{ otherwise.}
					\end{cases}
				\end{equation*}
				\changetwofour{For $p, p' \in \terminals \cup \{\emptyword, \bot \}$, we say $p$ and $p'$ are \emph{in conflict} if $p \neq \bot$ and $p' \neq \bot$, and $p \neq p'$. Let $\fcdatalogrule$ and $\anotherfcdatalogrule$ be linear $\fcdatalog$ rules \changefour{with the same head relation symbol}, with respective profiles $pro_{\fcdatalogrule}$ and $pro_{\anotherfcdatalogrule}$. Let $\topvar(\fcdatalogrule) = (x_1, \ldots, x_k)$. Note that $\topvar(\fcdatalogrule) =  \topvar(\anotherfcdatalogrule)$ due to both rules having the same head relation symbol. We say $pro_{\fcdatalogrule}$ and $pro_{\anotherfcdatalogrule}$ are \emph{in conflict} if there exists some $x \in \topvar(\fcdatalogrule)$ where $pro_{\fcdatalogrule}(x)$ is in conflict with $pro_{\anotherfcdatalogrule}(x)$}. 
			\end{defi}

			The function $pro_{\fcdatalogrule}$ represents how each $x \in \topvar(\fcdatalogrule)$ is processed in $\fcdatalogrule$. Let $H$ be the head relation symbol of $\fcdatalogrule$. In $\olla$ $\fcdatalog$, all we need to consider is either the leftmost or rightmost letter of each $\sub(x)$ for every substitution~$\sub$. We consider the leftmost letter if for rules with  head relation symbol $H$, we have that $x$ occurs in left-$\olla$ pattern equations, and analogously consider the rightmost letter if $x$ occurs in right-$\olla$ pattern equations.
			
			\begin{lem}
			\label{lem:profiles-deterministic}
			\changetwo{A guarded $\olla$ $\fcdatalog$ program $\fcdatalogprogram$ is globally deterministic if and only if for every relation symbol $R \in \relationsymbols$, for every pair of distinct rules $\fcdatalogrule, \anotherfcdatalogrule \in \fcdatalogruleset_{R}$, we have that $pro_{\fcdatalogrule}$ is in conflict with $pro_{\anotherfcdatalogrule}$.}
			\end{lem}
			\begin{proof}
				If there exists a pair of distinct rules $\fcdatalogrule, \anotherfcdatalogrule \in \fcdatalogruleset_{R}$ for some relation symbol $R \in \relationsymbols$ where $\we(\fcdatalogrule) \wedge \we(\anotherfcdatalogrule)$ is satisfiable, then there exists some top tuple $\vec{t_t}$ such that $(\vec{t_t}, \vec{t_1}) \in W_{\fcdatalogrule}$ and $(\vec{t}, \vec{t_2}) \in W_{\anotherfcdatalogrule}$, for some bottom tuples $\vec{t_1}, \vec{t_2}$. Hence $\fcdatalogprogram$ is globally nondeterministic. On the other hand, if $\we(\fcdatalogrule) \wedge \we(\anotherfcdatalogrule)$ is not satisfiable for all distinct $\fcdatalogrule, \anotherfcdatalogrule \in \fcdatalogruleset_{R}$, such a top tuple does not exist and hence $\fcdatalogprogram$ is globally deterministic.
				
				For the if direction, we assume for contradiction that $\fcdatalogprogram$ is not globally deterministic and for every pair of distinct rules $\fcdatalogrule, \anotherfcdatalogrule \in \fcdatalogruleset_{R}$, for every relation symbol~$R \in \relationsymbols$, we have that that $pro_{\fcdatalogrule}$ and $pro_{\anotherfcdatalogrule}$ are in conflict. As the profiles are in conflict, then for every pair~$\fcdatalogrule$~and~$\anotherfcdatalogrule$, there exists some $x \in \topvar(\fcdatalogrule)$ where $pro_{\fcdatalogrule}(x) = \mathsf{a}$ for $\mathsf{a} \in \terminals \cup \{\emptyword\}$ and $pro_{\anotherfcdatalogrule}(x) = \mathsf{b}$ for $\mathsf{b} \in (\terminals \cup \{\emptyword\})  \setminus \{\mathsf{a}\}$. 
	
				Therefore, we have two cases. In the first case we have $E_1 \coloneqq x \doteq y\mathsf{a}$ (or $E_1 \coloneqq x \doteq \mathsf{a}y)$ in $\we(\fcdatalogrule)$ and $E_2 \coloneqq x \doteq y\mathsf{b}$ (or $E_2 \coloneqq x \doteq \mathsf{b}y)$ in $\we(\anotherfcdatalogrule)$  for $\mathsf{a}, \mathsf{b} \in \terminals$ and $\mathsf{a} \neq \mathsf{b}$, and therefore $\we(\fcdatalogrule) \wedge \we(\anotherfcdatalogrule)$ is not satisfiable. In the second case, either $E_1$ or $E_2$ is replaced by $E_3 \coloneqq x \doteq \emptyword$, and we see again that $\we(\fcdatalogrule) \wedge \we(\anotherfcdatalogrule)$ is not satisfiable. As in each case we have that $\we(\fcdatalogrule) \wedge \we(\anotherfcdatalogrule)$ is not satisfiable, then for every pair of rules $\fcdatalogrule, \anotherfcdatalogrule \in \fcdatalogruleset_{R}$, we cannot have $(\vec{t_t}, \vec{t_1}) \in W_{\fcdatalogrule}$ and $(\vec{t_t}, \vec{t_2}) \in W_{\anotherfcdatalogrule}$, for some top tuple $\vec{t_t}$ and some bottom tuples $\vec{t_1}$ and  $\vec{t_2}$. Thus, we do have global determinism, and so have a contradiction.
				
				For the only if direction, we assume for contradiction that $\fcdatalogprogram$ is globally deterministic and there exists a pair of distinct rules $\fcdatalogrule, \anotherfcdatalogrule \in \fcdatalogruleset_{R}$ for some relation symbol $R \in \relationsymbols$ such that $pro_{\fcdatalogrule}$ and $pro_{\anotherfcdatalogrule}$ are not in conflict. As the profiles are not in conflict, then for all $x \in \topvar(\fcdatalogrule)$, we have that $pro_{\fcdatalogrule}(x)$ and $pro_{\anotherfcdatalogrule}(x)$ are not in conflict. Thus, for each $x \in \topvar(\fcdatalogrule)$, we have either $pro_{\fcdatalogrule}(x) = \bot$, or $pro_{\anotherfcdatalogrule}(x) = \bot$, or $pro_{\fcdatalogrule}(x) = pro_{\anotherfcdatalogrule}(x)$. Then $\we(\fcdatalogrule) \wedge \we(\anotherfcdatalogrule)$ is satisfiable and therefore there exists the top tuple $\vec{t_t}$ such that $(\vec{t_t}, \vec{t_1}) \in W_{\fcdatalogrule}$ and $(\vec{t_t}, \vec{t_2}) \in W_{\anotherfcdatalogrule}$, for some bottom tuples~$\vec{t_1}$ and $\vec{t_2}$. Global determinism is therefore violated, and thus we have a contradiction.
			\end{proof}

			Therefore, we can efficiently check if a given $\olla$ $\fcdatalog$ program is deterministic.
			
			\begin{prop}
			\label{lem:check-determinism-ptime}
			Local and global determinism for $\olla$ $\fcdatalog$ can be decided in polynomial time.
			\end{prop}
			\begin{proof}
			    \changetwo{Let $\fcdatalogprogram$ be an $\olla$ $\fcdatalog$ program with rule set $\fcdatalogruleset$. Let $m = \abs{\fcdatalogruleset}$ and let $p$ be the maximum number of pattern equations in a rule $\fcdatalogrule$, for all $\fcdatalogrule \in \fcdatalogruleset$. Let $t$ be the maximum $\abs{\topvar(\fcdatalogrule)}$ and let $v$ be the maximum $\abs{\topvar(\fcdatalogrule) \cup \bottomvar(\fcdatalogrule)}$ for all $\fcdatalogrule \in \fcdatalogruleset$.}
			    	
			    \changetwo{We first check every rule $\fcdatalogrule$ for local determinism by checking that all $x \in \topvar(\fcdatalogrule)$ appear in some pattern equation $E$ in $\we(\fcdatalogrule)$ or $x \in \bottomvar(\fcdatalogrule)$, and all $y \in \bottomvar(\fcdatalogrule)$ appear in some pattern equation $E$ in $\we(\fcdatalogrule)$ or $y \in \topvar(\fcdatalogrule)$. If so, then there is a partial function between the top and bottom variables. We can compute this in $\bigo(mpv)$.}
			    
			    \changetwo{We then verify global determinism. We first check if the program is guarded, which we can do in $\bigo(mp)$. Then, from Lemma \ref{lem:profiles-deterministic} we can check global determinism by checking that for every $R \in \relationsymbols$, for each pair of rules $\fcdatalogrule, \anotherfcdatalogrule \in \fcdatalogruleset_{R}$ where $\fcdatalogrule \neq \anotherfcdatalogrule$, we have that $pro_{\fcdatalogrule}$ is in conflict with $pro_{\anotherfcdatalogrule}$.  We can compute the profile for every rule in $\bigo(mpt)$. In the worst case, checking that $pro_{\fcdatalogrule}$ is in conflict with $pro_{\anotherfcdatalogrule}$ takes $\bigo(m^{2}t)$.}
			\end{proof}

			We thus have a fragment where we have reduced data complexity to $\logspace$, and without losing efficient checking of membership in the fragment. \changetwo{However,} the combined complexity for $\dolla$ $\fcdatalog$ is the same as for linear $\fcdatalog$.

			\begin{cor}
				\label{lem:olla-combined-complexity}
				Combined complexity of $\dolla$ $\fcdatalog$ is $\pspace$-complete.
			\end{cor}
			\begin{proof}
			The $\pspace$ upper bound follows directly from Lemma \ref{cor:linear-combined-membership-pspace}. For $\pspace$-hardness, let $\fcdatalogprogram$ be the program defined using the construction in the proof of Lemma~\ref{lem:linear-combined-pspace-hard}. Observe that $\fcdatalogprogram$ is both an $\olla$ $\fcdatalog$ program and deterministic.
			\end{proof}

			We have therefore defined the endpoints of an infinite range of fragments that all capture $\logspace$. At one end, $\dolla$ $\fcdatalog$ has just enough syntax to simulate multi-headed two-way deterministic finite automata, but has easy checking of determinism. At the other end, deterministic linear $\fcdatalog$ has a richer syntax but has more expensive checking of determinism. From this range, there is therefore the opportunity to extend $\dolla$ $\fcdatalog$ to make writing programs more natural and convenient. We shall demonstrate such a situation in Section~\ref{sec:application} when designing an appropriate fragment to model deterministic regex. However, even the most restrictive fragment in this range does not improve combined complexity any more than linear $\fcdatalog$ does, and this motivates Section~\ref{subsec:sd}. 

		\subsection{Strictly Decreasing}
			\label{subsec:sd}
			\changeother{Our next goal is more efficient combined complexity.
			We introduce a further restriction called \emph{strictly decreasing}, and show that this leads to  linear combined complexity.}

			\begin{defi}
				\label{def:sd}
				Let $\fcdatalogprogram$ be a $\dolla$ $\fcdatalog$ program with rule set $\fcdatalogruleset$, and let $\fcdatalogruleset' \subseteq \fcdatalogruleset$ be the rules containing both: a relation atom with a relation symbol that is mutually recursive with the head relation symbol, and at least one pattern equation. Let $w$ be the input word. We say $\fcdatalogprogram$ is \emph{strictly decreasing} ($\sd$) if for every rule~$\fcdatalogrule \in \fcdatalogruleset'$ and its head relation symbol $H$:
				\begin{enumerate}
					\item \label{condition1} there exist $x \in \topvar(\fcdatalogrule)$ and $y \in \bottomvar(\fcdatalogrule)$ such that $y$ occurs in a pattern equation with~$x$, and for all substitutions $\sub$ such that $(w, \sub) \models \fcdatalogrule$ we have $\abs{\sub(y)} < \abs{\sub(x)}$, and
					\item for the relation symbol $M$ in the body of $\fcdatalogrule$ that is mutually recursive with $H$, for every rule~$\anotherfcdatalogrule \in \fcdatalogruleset'$ with head relation symbol $M$, condition \eqref{condition1} holds for some $x' \in \topvar(\anotherfcdatalogrule)$ and some $y' \in \bottomvar(\anotherfcdatalogrule)$, and $\sub(y) = \sub'(x')$ for all substitutions $\sub$ such that $(w, \sub) \models \fcdatalogrule$ and all substitutions $\sub'$ such that $(w, \sub') \models \anotherfcdatalogrule$.
				\end{enumerate}
			\end{defi}

			We now see that this restriction significantly improves the combined complexity. In fact, if the maximum relation symbol arity $k$ is fixed (or assumed to be much smaller than $\abs{w}$), then this is linear. Furthermore, the preprocessing is not dependent on the word $w$.
			
			\begin{thm}
				\label{thm:sd-comb-compl}
				Given a word $w$ and an $\sddolla$ $\fcdatalog$ program $\fcdatalogprogram$ where the cardinality of the relation symbol set $\abs{\relationsymbols} = n$, and the maximum relation symbol arity is $k$, we can decide $w \in \lang(\fcdatalogprogram)$ in $\bigo(\abs{w}k)$ time after $\bigo(n\abs{\terminals})$ preprocessing.
			\end{thm}
			\begin{proof}
				The preprocessing step is to build a lookup table for which rule to apply. For each relation symbol $R$ there is at most $\abs{\terminals}$ rules where $R$ is in the head, and so we have $\bigo(n\abs{\terminals})$ preprocessing. In a strictly decreasing $\dolla$ $\fcdatalog$ program, as we constantly reduce the length of one variable, the number of rule applications is bounded by $\abs{w}$. We then use the lookup table to decide which rule to apply, and as a rule can have at most $\bigo(k)$ pattern equations, in total we use $\bigo{(\abs{w}k)}$ time.
			\end{proof}

			Therefore, although each of our range of fragments that all capture $\logspace$ do not reduce combined complexity any more than \changetwo{is the case for linear $\fcdatalog$}, when we add the extra dimension of strictly decreasing, we can reduce the combined complexity to linear time.
		
		\subsection{FC-Datalog as Generalized Automata}
			\label{subsec:comparison}
			Throughout Section \ref{sec:efficient} and the current section, we have defined a series of fragments of $\fcdatalog$. Furthermore, as discussed in Section \ref{subsec:olla}, we have in fact defined the endpoints of an infinite range of fragments that all capture $\logspace$; at one end we have $\dolla$ $\fcdatalog$, which has efficient determinism checking but severely restricted syntax, and the other we have deterministic linear $\fcdatalog$, which has a richer syntax but expensive checking of determinism. These fragments capture $\logspace$ as they are able to simulate multi-headed two-way deterministic finite automata, and thus we can view our $\fcdatalog$ fragments as generalizations of this automata model, or its nondeterministic counterpart if the fragment is not deterministic.

			As $\dolla$ $\fcdatalog$, by design, permits just enough to simulate these automata, we can naturally see a $\dolla$ $\fcdatalog$ program as a generalization of multi-headed two-way DFA: every relation acts as a state, every rule acts as a transition, and every string variable acts a head which can be moved, added and deleted (as the number of heads can change per state). In each transition we read one letter and move, add, or delete the heads accordingly. 
			
			The other fragments of $\fcdatalog$ discussed in this paper can be seen as further generalizations. For example, Section \ref{sec:application} introduces $\dollaplus$ $\fcdatalog$ which additionally allows pattern equations such as $x \doteq yy$. Instead of just moving a head by one letter, this is an operation that either splits a memory in half or doubles it. We can then see a $\dollaplus$ $\fcdatalog$ program as an generalized multi-headed two-way DFA that can perform nonregular string computations in the transitions and where heads can read words rather than letters. Further fragments from our infinite range that permit other syntactic additions could then also be seen as further generalizations.
			
			Akin to how we can see a $\dolla$ $\fcdatalog$ program as a generalized multi-headed two-way DFA, we can see an $\olla$ $\fcdatalog$ program as a generalized multi-headed two-way NFA. Then, as our syntax grows, the connection to traditional automata gradually loosens. We can see a linear $\fcdatalog$ program as a further generalization of a multi-headed two-way NFA, where reading letters at head positions is replaced with a declarative programming language that checks if the rule applies and computes the operations. A deterministic linear $\fcdatalog$ program is an analogous generalization of a multi-headed two-way DFA. The increasing gap to traditional automata is reflected in the fact that determinism is now expensive to verify. 
			
			Finally, for unrestricted $\fcdatalog$, we can understand every relation symbol in the body of a rule as a call of a subroutine,  which splits the automaton into parallel copies, each of which must terminate. At this general level, the gap to traditional automata is so large that the connection is no longer natural.
		
	\section{Applying the Framework: Simulating DRX}
		\label{sec:application}
		In this section, we tailor a fragment from the \changetwo{range} we defined in Section \ref{sec:framework} to a specific application. We apply our model to simulate deterministic regex, introduced in~\cite{deterministicRegex}.
		
		As in \cite{deterministicRegex}, we define regular expressions as usual and extend these with back-references to define \emph{regex}. For $x \in \vars$ and a regular expression $\delta$, we thus add the expressions $x$ and $\langle x \colon \delta \rangle$ to our syntax. The expression $\langle x \colon \delta \rangle$ matches $\delta$ and saves the string that is matched by $\delta$ in the  \emph{memory} $x$. All further occurrences of $x$ are \emph{recalls} of memory~$x$, which we match using the content of the variable that was saved earlier. For readability, we use $\cdot$ for concatenation.

		\begin{exa}
			\label{ex:det-regex}
			Let $\deterministicregex \coloneqq \langle x \colon (\mathsf{a} \vee \mathsf{b})^{+} \rangle \cdot \mathsf{d} \cdot x$. Then $\deterministicregex$ matches all words $u\mathsf{d}u$ where $u \in \{\mathsf{a}, \mathsf{b}\}^{+}$.
		\end{exa}
		
		Every regular expression can be converted into a finite automaton using the classical Glushkov construction from \cite{Glushkov}. If the result of this construction is deterministic, then the regular expression is deterministic. Deterministic regular expressions define a strict subclass of the regular languages, and have a more efficient membership problem than general regular expressions. 
		For a deterministic regular expression $\deterministicregex$ and a word $w$, we can decide membership in $\bigo(\abs{\terminals} \abs{\deterministicregex} + \abs{w}$) (see \cite{BruggermanKlein, Ponty}) or $\bigo(\abs{\deterministicregex} + \abs{w} \log \log \abs{\deterministicregex})$ (see \cite{DeterministicRegularExpressions}).

		In \cite{deterministicRegex}, regex were combined with deterministic regular expressions to define \emph{deterministic regex}, which can define nonregular languages, and also have an efficient membership problem. Membership can be decided in $\bigo(\abs{\terminals}\abs{\deterministicregex}n + k \abs{w})$ for a word $w$ and a deterministic regex $\deterministicregex$ with \changethree{ $k$ distinct variables and $n$ total occurrences} of terminal symbols or variable references (see Theorem 5 of \cite{deterministicRegex}). Similarly to the Glushkov construction, every regex~$\deterministicregex$ has a corresponding \emph{memory finite automaton with trap state} (\emph{$\dtmfa$}) $M_{\deterministicregex}$ \changetwo{(where the trap state handles memory recall failures)}, and $\deterministicregex$ is deterministic if $M_{\deterministicregex}$ is deterministic (see~\cite{deterministicRegex} for details). We call the class of all deterministic regex $\drx$. 

		\begin{exa}  Let $\deterministicregex$ be the regex defined in Example \ref{ex:det-regex}, and let $\deterministicregex' \coloneqq \langle x \colon (\mathsf{a} \vee \mathsf{b})^{\ast} \rangle \cdot x$. Then $\lang(\deterministicregex) \coloneqq \{u\mathsf{d}u \mid u \in \{\mathsf{a}, \mathsf{b}\}^{\ast}\}$ and $\lang(\deterministicregex') \coloneqq \{uu \mid u \in \{\mathsf{a}, \mathsf{b}\}^{\ast}\}$. Then $\deterministicregex \in \drx$ and $\deterministicregex' \notin \drx$, as $M_{\deterministicregex}$ is deterministic and $M_{\deterministicregex'}$ is not deterministic. Intuitively, this is because $\deterministicregex$ has only one choice of when to stop matching the first $u$, whereas $\deterministicregex'$ does not.
		\end{exa}

		As in \cite{deterministicRegex}, we can assume w.l.o.g. that a regex does not recall empty memories, does not start to save into a memory that is already being saved to, and does not reset a memory to its initial value. We first show that \changeother{for any $\deterministicregex\in\drx$, we can express $\lang(\deterministicregex)$ in $\sddolla$ $\fcdatalog$}, but with a large number of rules. \changethree{We can parameterize such a program using the number of memories, terminal symbols and memory recalls in $\deterministicregex$}. Here we use \changetwo{$\emptyword$-semantics\footnote{In $\emptyword$-semantics we treat everything not initialized as $\emptyword$ \changetwo{(see e.g. Section 8.2.1 in \cite{deterministicRegex}).}}.}

		\begin{thm}
			\label{thm:dolla-regex}
			Let $\deterministicregex \in \drx$ have $k$ memories. \changeother{Let the total number of terminal symbols and memory recalls in $\deterministicregex$ be $n$}. We can express $\lang(\deterministicregex)$ with an $\sddolla$ $\fcdatalog$ program~$\fcdatalogprogram$ that has at most \changeother{$k + n + 2$} relation symbols and at most $k(\abs{\terminals} + 1) + n(n + \changeother{3}) + 1$ rules.
		\end{thm}
		\begin{proof}
			We define an explicit construction from a deterministic regular expression with back-references $\deterministicregex$ to a $\dolla$ $\fcdatalog$ program $\fcdatalogprogram$. Let $t$ be the number of terminal symbols and back-references in $\deterministicregex$. Let $\deterministicregex[i]$ be the terminal symbol or back-reference at index $i$ in $\deterministicregex$.  We start with $\relationsymbols = \{\ans, Q_0\}$. We then, for all $1 \leq i \leq t$, add a new relation symbol $Q_i$ to~$\relationsymbols$. We add the initial rule $\ans() \shortleftarrow Q_0(x', x_1, \ldots, x_{k}) \shortleftarrow x' \doteq \inputsymbol \rc x_1 \doteq \emptyword \rc \ldots \rc x_{k} \doteq \emptyword$. We can model processing a terminal $\mathsf{a}$ using $\fcdatalog$ rules of the form:
			\begin{equation*}
				Q_{\text{src}}(u, x'_1, \ldots x'_k) \shortleftarrow Q_{\text{dst}}(v, x_1, \ldots x_k)\rc u \doteq \mathsf{a}v \rc x_1 \doteq x'_1\mathsf{a} \rc \ldots \rc x_k \doteq x'_k\mathsf{a}.
			\end{equation*}
			The pattern equation $u \doteq \mathsf{a}v$ consumes the next letter and $v$ holds the rest of the word to be processed. The pattern equations of the form $x_m \doteq x'_m\mathsf{a}$ adds the letter to the appropriate variable; if the letter does not need to be added to this variable we can omit the pattern equation entirely, and rename $x'_m$ to just $x_m$. For all letters $\deterministicregex[i]$ that can be the first letter, let src be $0$ and dst be $i$. For all letters $\deterministicregex[j]$ and for all successor letters $\deterministicregex[i]$ to $\deterministicregex[j]$, let src be $j$ and dst be $i$.
			
			Modelling the processing of a back-reference $x'_n$ requires more rules. To do this, we first use rules of the form:
			\begin{equation*}
				Q_{\text{src}}(u, x'_1, \ldots x'_k) \shortleftarrow Q_{\text{dst}}(v, x_1, \ldots x_k)\rc R_\ell(u, v, x_n, x'_1, x_1\ldots, x'_{\ell}, x_{\ell}).
			\end{equation*}
			The values for src and dst are applied in the same way as for terminals and the extra relation~$R_\ell$ can be solved as a subroutine using the following rules. We have for every terminal $\mathsf{a} \in \terminals$ and every $ 0 \leq \ell \leq k-1$ the rules:
			\begin{multline*}
				R_\ell(u, v, x_n, x'_1, x_1, \ldots, x'_{\ell}, x_{\ell}) \shortleftarrow x_n \doteq \mathsf{a}x'_n \rc u \doteq \mathsf{a}u' \rc
				\\ x''_1 \doteq x'_1{\mathsf{a}} \rc \ldots \rc x''_{\ell} \doteq x'_{\ell}\mathsf{a} \rc  R_\ell(u', v, x'_n, x''_1, x_1, \ldots, x''_{\ell}, x_{\ell});
			\end{multline*}
			and the rule:
			\begin{equation*}
				R_\ell(u, v, x'_1, x_1, \ldots, x'_{\ell}, x_{\ell}) \shortleftarrow x_n \doteq \emptyword \rc u \doteq v \rc x'_1 \doteq x_1 \rc \ldots \rc x'_{\ell} \doteq x_{\ell}.
			\end{equation*}
			Here $R_{\ell}$ computes adding each letter of the saved variable's value, and writes to $\ell$ other variables. Finally, for all symbols $\deterministicregex[i]$ that can be read last and the word be accepted, we add the rule $Q_i(u, x_1, \ldots, x_k) \shortleftarrow u \doteq \emptyword$
		\end{proof}

		The program $\fcdatalogprogram$ that is the result of our construction requires a rule for each pair $(\mathsf{a}, \ell)$, for $\mathsf{a} \in \terminals$ and $0 \leq \ell \leq k-1$. 
		This is because the syntax permits matching only one letter at a time, and so the program requires many rules to read memories letter by letter. 
		We can achieve this with a much more concise program if we relax the syntax slightly, to what we call $\dollaplus$ $\fcdatalog$.
		This sits between $\dolla$ $\fcdatalog$ and deterministic linear $\fcdatalog$, and it retains the polynomial time checking of determinism. 
		To define this, we introduce the concept of a symbol being \emph{uniquely defined} for top-down evaluation. 

		\begin{defi}
		Let $\fcdatalogrule$ be an $\fcdatalog$ rule. We inductively define \emph{uniquely defined} symbols: As base rules, the universe variable $\inputsymbol$, every $\mathsf{a} \in \terminals$ and every  $x \in \topvar(\fcdatalogrule)$ are uniquely defined. Then, if we have some pattern equation~$\atom$ where $\text{\Var}(\atom) = \{x_1, \ldots x_k\}$ and exactly one variable $x_i$ is not uniquely defined, then $x_i$ becomes uniquely defined.
		\end{defi}
		
		Example \ref{ex:top-down-eval} illustrates how top variables are uniquely defined for top-down evaluation.
		
		\begin{exa}
			\label{ex:unq-def}
			Let $\fcdatalogrule = R_1(x_1, x_2) \shortleftarrow  \atom_1, \atom_2, R_2(y_1, y_2)$ be a linear $\fcdatalog$ rule where $\atom_1 \coloneqq x_1 \doteq x_2y_1$ and $\atom_2 \coloneqq y_1 \doteq y_2y_2$. From the base rules, $x_1$ and $x_2$ are uniquely defined. In the first iteration, as all $ z \in \Var(\atom_1) \setminus \{y_1\}$ are uniquely defined, $y_1$ is uniquely defined. In the second iteration, as all $ z \in \Var(\atom_2) \setminus \{y_2\}$ are uniquely defined, $y_2$ is uniquely defined.
		\end{exa}

		\changetwo{In deterministic linear and $\dolla$ $\fcdatalog$, by Definition \ref{def:determinism-criteria}, we ensure local determinism by requiring the relation $W_{\fcdatalogrule}$ to be a partial function} for every rule $\fcdatalogrule$, implicitly requiring every variable to be uniquely defined. We ensure local determinism in $\dollaplus$ $\fcdatalog$ by reasoning directly over uniquely defined variables.
		
		\begin{defi}
		\label{def:dollaplus}
			Let $\fcdatalogprogram$ be a linear $\fcdatalog$ program with rule set $\fcdatalogruleset$. We say $\fcdatalogprogram$ is a $\dollaplus$ $\fcdatalog$ program if it is globally deterministic and for every rule $\fcdatalogrule \in \fcdatalogruleset$, \changetwo{every variable that occurs in $\fcdatalogrule$ is uniquely defined}.
		\end{defi}
		
		\begin{exa}
			The $\fcdatalog$ program $\fcdatalogprogram$ given in Example \ref{ex:fc-datalog-program} that is not a $\dolla$ $\fcdatalog$ program (see Example \ref{ex:not-dolla}) is in fact a $\dollaplus$ $\fcdatalog$ program.
		\end{exa}

		\changetwofour{We define $\sd$ for $\dollaplus$  the same way as for $\dolla$   (Definition~\ref{def:sd}).  $\sddollaplus$ $\fcdatalog$ generalizes $\sddolla$ $\fcdatalog$. We now show that despite being extensions, $\dollaplus$ $\fcdatalog$ retains the low data complexity (Theorem~\ref{thm:dollaplus-complexity}) and the efficient checking of determinism (Proposition \ref{prop:dollaplus-checkable}) of $\dolla$ $\fcdatalog$, and $\sddollaplus$ $\fcdatalog$ retains the low combined complexity of $\sddolla$ $\fcdatalog$ (Theorem~\ref{thm:dollaplus-complexity})}. 
		
		\begin{thm}
			\label{thm:dollaplus-complexity}
			$\dollaplus$ $\fcdatalog$ captures $\logspace$, and given a word $w$ and an $\sddollaplus$ $\fcdatalog$ program $\fcdatalogprogram$ with $n$ relation symbols and maximum relation symbol arity $k$, we can decide $w \in \lang(\fcdatalogprogram)$ in $\bigo(\abs{w}k)$ time after $\bigo(n\abs{\terminals})$ preprocessing.
		\end{thm}
		\begin{proof}
			For data complexity, a $\logspace$ upper bound follows from Theorem~\ref{thm:dolla-capture-logspace}, as we ensure global determinism in the same way and ensure local determinism by permitting only uniquely defined variables. Then, expressing all of $\logspace$ follows directly from Lemma~\ref{lem:dolla-all-logspace}. For combined complexity, this follows directly from Theorem \ref{thm:sd-comb-compl}.
		\end{proof}

		\begin{prop}
			\label{prop:dollaplus-checkable}
			Membership of $\dollaplus$ $\fcdatalog$  can be decided in polynomial time.
		\end{prop}
		\begin{proof}
			We can check local determinism by verifying that each variable in each rule is uniquely defined. As this is a syntactic criterion, we can check this in polynomial time. Checking global determinism can be done in the same way as for Proposition \ref{lem:check-determinism-ptime}.
		\end{proof}
		
		\changetwo{Where a $\dolla$ $\fcdatalog$ program is required to match} letter-by-letter, this is no longer the case for $\dollaplus$ $\fcdatalog$, and a consequence of this is that we can simulate a deterministic regex with a much more concise $\sddollaplus$ $\fcdatalog$ program.

		\begin{thm}
			\label{thm:dollaplus-regex}
			Let $\deterministicregex \in \drx$ have $k$ memories. \changeother{Let the total number of terminal symbols and memory recalls in $\deterministicregex$ be $n$}. We can express $\lang(\deterministicregex)$ with an $\sddollaplus$ $\fcdatalog$ program~$\fcdatalogprogram$ that has at most $n + 2$ relation symbols and at most $n(n + \changeother{3}) + 1$ rules.
		\end{thm}
		\begin{proof}
			We give an adaptation to the construction in Theorem \ref{thm:dolla-regex}. As we can use further forms of pattern equations in $\dollaplus$ $\fcdatalog$, we can simulate the processing of back-references in the same way as we process terminals in $\dolla$ $\fcdatalog$. To model processing the back-reference $x'_n$, we use the rule:
			\begin{equation*}
				Q_{\text{src}}(u, x'_1, \ldots, x'_k) \shortleftarrow Q_{\text{dst}}(v, x_1, \ldots, x_k)\rc u \doteq x'_{n}v \rc x_{1} \doteq x'_{1}x'_n \rc \ldots \rc x_{\ell} \doteq x'_{\ell}x'_{n}.
			\end{equation*}
			Again, if we do not write to the variable $x'_m$, we can omit the equation $x_m \doteq x'_{m}x'_{n}$ and rename $x'_m$ in the head to $x_{m}$. We therefore eliminate having a rule for each pair $(\mathsf{a}, \ell)$ and a final rule for each $\ell$, for $0 \leq \ell \leq k-1$. Therefore we reduce the number of rules by $k(\abs{\terminals} + 1)$.
		\end{proof}
		
		\changeall{As the following example shows, the extra flexibility permitted in the syntax of $\dollaplus$ $\fcdatalog$ allows us to write programs much more conveniently than for $\dolla$ $\fcdatalog$}.

		\begin{exa}
		\changeall{
			In $\dollaplus$ $\fcdatalog$, we can model processing a memory $x'_n$ with a single rule:
			\begin{align*}
				Q'(u, x'_1, \ldots, x'_k) \shortleftarrow Q(v, x_1, \ldots, x_k) \rc u \doteq x'_{n}v \rc x_1 \doteq x'_1x'_{n} \rc \ldots \rc x_k \doteq x'_kx'_{n}.
			\end{align*}
			In $\dolla$ $\fcdatalog$, to model processing a memory $x'_n$, we use a new relation symbol $R_\ell$ and the rules:
			\begin{align*}					
				Q'(u, x'_1, \ldots, x'_k) &\shortleftarrow Q(v, x_1, \ldots, x_k) \rc R_{\ell}(u, v, x_n, x'_1, x_1, \ldots, x'{\ell}, x_{\ell});
				\\R_\ell(u, v, x'_1, x_1, \ldots, x'_{\ell}, x_{\ell}) &\shortleftarrow x_n \doteq \emptyword \rc u \doteq v \rc x'_1 \doteq x_1 \rc \ldots \rc x'_{\ell} \doteq x_{\ell};
			\end{align*}
			and for every $\mathsf{a} \in \terminals$ a rule:
			\begin{multline*}
				R_\ell(u, v, x_n, x'_1, x_1, \ldots, x'_{\ell}, x_{\ell}) \shortleftarrow x_n \doteq \mathsf{a}x'_n \rc u \doteq \mathsf{a}u' \rc
				\\ x''_1 \doteq x'_1{\mathsf{a}} \rc \ldots \rc x''_{\ell} \doteq x'_{\ell}\mathsf{a} \rc  R_\ell(u', v, x'_n, x''_1, x_1, \ldots, x''_{\ell}, x_{\ell}).
			\end{multline*}}
		\end{exa}

		We have thus demonstrated the design of a tailored fragment in our \changetwo{range} spanned by $\dolla$ and deterministic linear $\fcdatalog$, in this case by adding pattern equations with uniquely defined variables.  \changefour{The $\dollaplus$ and $\sddollaplus$ fragments retain the desirable properties of the restrictive $\dolla$ and $\sddolla$ fragments. As discussed in Section~\ref{subsec:comparison}, 
			we can see $\dollaplus$ $\fcdatalog$ as a generalization of multi-headed two-way DFAs which is then not restricted to left-to-right parsing. As the next example shows, it can express context-free languages  which cannot be recognized by a DPDA (see e.g. Chapter 6 of~\cite{cinderella})}.
		
		\begin{exa}
		\label{exmp:palindrome}
		The palindrome language can be expressed deterministically by the $\dollaplus$ $\fcdatalog$ program that has the rules $\ans() \shortleftarrow R(\inputsymbol)$; $R(x) \shortleftarrow x \doteq \emptyword$ and for each $\mathsf{a} \in \terminals$, the rules $R(x) \shortleftarrow x \doteq \mathsf{a}y\mathsf{a}\smallrc R(y)$; $R(x) \shortleftarrow x \doteq \mathsf{a}$.
		 This evaluates in $\ceil{\abs{w}/2} + 1$ steps, which is linear.
		\end{exa}
		
		Furthermore, as we can simulate $\drx$, another feature that can be added to these tailored fragments are atoms that match $\drx$, which we can solve as a subroutine. We say a \emph{$\drx$-constraint} is an expression $(x \dot{\in} \deterministicregex)$ for $x \in \vars$ and a deterministic regex $\deterministicregex$, that denotes $x$ is mapped to an element $u \in \lang(\deterministicregex)$ and $u \factor w$, where $w$ is the input word. 

		To guarantee we remain in $\logspace$, we must ensure programs remain locally and globally deterministic. As $\drx$-constrains only check if a variable matches a deterministic regex, we can decide local determinism as we would without the $\drx$-constraints. To retain global determinism (Definition \ref{def:determinism-criteria}), if using deterministic regexes $\deterministicregex_1, \ldots, \deterministicregex_n$ in multiple rules with the same head relation symbol, we must ensure that always at most one rule can accept. We thus need to decide if $\bigcap_{i=1}^{n} \lang(\deterministicregex_i) = \emptyset$. 
		Unfortunately, intersection-emptiness  for $\drx$ is undecidable (see Theorem 9 of \cite{deterministicRegex}). We say $\deterministicregex \ \in \drx$ is variable-star-free if each of its sub-regexes under a Kleene-star or Kleene-plus do not contain any variable operations. \changeother{I}ntersection-emptiness for variable-star-free $\drx$ is at least word equations-hard (see Proposition 8 of \cite{deterministicRegex}). Thus, adding $\drx$-constraints means we  lose efficient checking of determinism, another example of the trade-off between richer syntax and efficient determinism checking.
		
		We can retain efficient determinism checking if we limit where we include $\drx$-constraints. If $\abs{\fcdatalogruleset_{R}} = 1$ for some relation symbol $R$, global determinism for $\fcdatalogruleset_{R}$ is inherent. \changeother{If we limit} inclusions of $\drx$-constraints to rules \changeother{$\fcdatalogrule \in \fcdatalogruleset_{R}$ where $\abs{\fcdatalogruleset_{R}} = 1$, we can verify} determinism by verifying every $\drx$-constraint is in such rules. We can decide this in polynomial time. Combining this with Proposition~\ref{prop:dollaplus-checkable}, we can check the whole program's determinism in polynomial time.

		\begin{exa}
		The $\fcdatalog$ program $\fcdatalogprogram = (\inputsymbol, \relationsymbols, \fcdatalogruleset)$ where $\fcdatalogruleset$ is:
		\begin{align*}
			\ans() &\shortleftarrow R_1(\inputsymbol); &
			R_1(x) &\shortleftarrow x  \doteq \mathsf{a}y\mathsf{c}\rc R_1(y); \\
			R_2(x) &\shortleftarrow (x \dot{\in} \langle y \colon (\mathsf{c} \vee \mathsf{d})^{+} \rangle \cdot \mathsf{a} \cdot y); &
			R_1(x) &\shortleftarrow x  \doteq \mathsf{b}y\mathsf{b}\rc R_2(y).
		\end{align*}
		defines the language $\{\mathsf{a}^{z}\mathsf{b}w\mathsf{a}w\mathsf{b}\mathsf{c}^{z} \mid z \in \natnums, \inlineextraspace w \in \{ \mathsf{c}, \mathsf{d}\}^{\ast}\}$. Furthermore, for all $R \in \relationsymbols$  and corresponding subsets of rules $\fcdatalogruleset_{R}$, we have that $\fcdatalogruleset_{R_2}$ is the only one that has rules containing $\drx$-constraints, and $\abs{\fcdatalogruleset_{R_{2}}} = 1$.
		\end{exa}
		
		Finally, we show a comparison between $\dollaplus$ $\fcdatalog$ and $\fcreg$, the extension of $\fc$ with constraints that decide membership of regular languages, as defined in \cite{FC}.

		\begin{rem}
			$\fcreg$, which captures the expressive power of generalized core spanners, has $\logspace$ data complexity. However, $\fcreg$ does not capture $\logspace$, as it cannot express the language $\{a^{n}b^{n} \mid n \in \natnums\}$ (see \cite{FC}). We can naturally express this language in $\dollaplus$ $\fcdatalog$ as the following example shows.
		\end{rem}
		
		\begin{exa}
		We can express $\{a^{n}b^{n} \mid n \in \natnums\}$ with the $\dollaplus$ $\fcdatalog$ program that has the rules: $\ans() \shortleftarrow R(\inputsymbol);$ $R(x) \shortleftarrow x \doteq \emptyword;$ and $R(x) \shortleftarrow x \doteq \mathsf{a}y\mathsf{b} \smallrc R(y)$.
		\end{exa}
		
	\section{Conclusions and Future Work}
		\label{sec:conclusions}
		The logic $\fc$ was proposed in \cite{FC} with the original motivation of capturing classes of spanners, a well-studied tool for information extraction. That said, independent of the spanner connection, $\fc$ is also as a general framework for logic on strings. To this end, $\fcdatalog$ was proposed in \cite{FC} as an extension of the logic $\fc$ with recursion, which captures the complexity class $\p$. We can thus view $\fcdatalog$ as a version of $\datalog$ on strings with a decidable model checking problem, where the same problem for a previous approach towards $\datalog$ on strings is undecidable. We could also view it as a language for expressing relations that can be used in spanner selections. We first showed that combined complexity of $\fcdatalog$ is $\exptime$-complete, presenting an opportunity for optimization of model checking. In this paper we identified fragments of $\fcdatalog$ with an efficient model checking problem, for both data and combined complexity, by performing an analysis of four restrictions: linearity, determinism, one letter lookahead and strictly decreasing.

		In Section \ref{subsec:linearity} we proposed linear $\fcdatalog$, which captures $\nlogspace$.  Then in Section~\ref{subsec:determinism} we identified and eliminated nondeterminism. Thus, deterministic linear $\fcdatalog$ captures $\logspace$. However, checking membership in this fragment is expensive.
		
		In Section \ref{subsec:olla} we imposed the one~letter lookahead restriction on the pattern equations in a program to obtain deterministic $\olla$ ($\dolla$) $\fcdatalog$, which has both desirable properties: it captures $\logspace$, and membership in the fragment can be checked in polynomial time. We also showed that combined complexity for both linear and $\dolla$ $\fcdatalog$ is $\pspace$-complete. We thus added the further restriction of strictly decreasing~($\sd$) in Section \ref{subsec:sd}, and showed that $\sddolla$ $\fcdatalog$ has linear combined complexity. 
		
		We hence established the endpoints of a range of $\fcdatalog$ fragments that all capture $\logspace$, and showed how we can restrict this further to reduce combined complexity to linear time. We have therefore paved the way to construct further fragments which can be tailored for particular applications. We illustrated tailoring a fragment in Section \ref{sec:application}, where we constructed  $\dollaplus$ $\fcdatalog$. This allows us to straightforwardly model deterministic regex without the need for a technically involved automata model. Yet, as for $\dolla$ $\fcdatalog$: we capture $\logspace$, membership in the fragment can be checked in polynomial time, and the strictly decreasing programs have linear combined complexity. We then showed how deterministic regex can be used as atoms, and how we can restrict where they are used so as to not compromise on these desirable properties.
		
		Where we added additional forms of pattern equations in our example tailored fragment, there are various other syntactic additions that we could consider. As $\logspace$ is closed under complement, we could add stratified negation without affecting the combined complexity. Other convenient additions could include predicates that express properties of a string such as ``is a letter", and functions that build on these to express properties such as ``is the first/last letter". We could also add regular constraints: atoms that work in the same way as $\drx$-constraints, but match a classical regular expression rather than a deterministic regex, as in $\fcreg$. Using these or other syntactic additions, we could further map the range between $\dolla$ and deterministic linear $\fcdatalog$, and investigate how these additions affect checking membership in the fragment. Furthermore, we could also look to find sufficient conditions to ensure membership can be easily checked.
		
		As demonstrated in \cite{SplittingAtoms}, there are structure criteria such as acyclicity for conjunctive queries in $\fc$ ($\fccq$) that improve efficiency. As we mentioned in Section \ref{sec:preliminaries}, every rule of an $\fcdatalog$ program can be seen as a conjunctive query in $\fc$. We could therefore add such structure criteria to $\fcdatalog$ rules and examine how this improves the complexity of $\fcdatalog$ fragments further. Another direction for research is to look at inexpressibility for these fragments. However, even for $\fc$ (as opposed to $\fcdatalog$), finding inexpressibility results is challenging (see \cite{EFGames}). Another logical next step is to use our restrictions for $\fcdatalog$ to identify natural fragments in the spanner setting with desirable properties.
		
		Finally, we can also see $\fcdatalog$ as a generalization of range concatenation grammars~($\rcg$s). See~\cite{rcgChapter} for details of $\rcg$s and see \cite{FC} for details on the generalization. Thus parsing techniques for these grammars such as those in \cite{Kallmeyer} could give rise to further efficient $\fcdatalog$ fragments. In the other direction, further results for $\fcdatalog$ could bring about further developments in the parsing of $\rcg$s.		

	\section*{Acknowledgement}
	\noindent
	The authors would like to thank the anonymous reviewers of the previous versions for their detailed and helpful feedback.
	
	\section*{Data Availability}
	\noindent
	No data was generated, captured or analyzed for this work.
		
	\bibliographystyle{alphaurl}
	\bibliography{references}
	
\end{document}